\documentclass[aps,pre,preprint,groupedaddress]{revtex4-1}
\usepackage{graphicx}
\usepackage{amsthm,amssymb,amsmath,cancel,dsfont}
\usepackage[hidelinks,pagebackref=false,pdfnewwindow=true]{hyperref}
\usepackage{bm}

\newcommand{\dd}{\partial}

\begin{document}


\title{Impact of lattice rotation on dislocation motion}

\author{Brent Perreault}
\author{Jorge Vi\~nals}

\affiliation{School of Physics and Astronomy, and Minnesota Supercomputing 
Institute, University of Minnesota, Minneapolis, Minnesota 55455, USA}

\author{Jeffrey M. Rickman}
\affiliation{Department of Materials Science and Engineering, Lehigh
  University, Bethlehem, Pennsylvania 18015, USA}
\date{\today}

\begin{abstract}

We introduce a phenomenological theory of dislocation motion
appropriate for two dimensional lattices. A coarse grained description is proposed
that involves as primitive variables local lattice rotation
and Burgers vector densities along distinguished slip systems of the lattice.
We then use symmetry considerations to propose phenomenological equations for
both defect energies and their dissipative motion. As a consequence,
the model includes explicit dependences on the local state of lattice
orientation, and allows for differential defect mobilities along
distinguished directions. Defect densities and lattice rotation need
to determined self consistently and we show specific results for both
square and hexagonal lattices. Within linear response, dissipative
equations of motion for the defect densities are derived which contain
defect mobilities that depend nonlocally on defect distribution. 

\end{abstract}

\maketitle

\section{Introduction}

A phenomenological model of dislocation motion in two dimensional lattices is
introduced which is based on a coarse grained Burgers vector density. We extend
existing treatments that are based on dissipative motion driven by
plastic free
energy minimization by introducing anisotropic mobilities along locally rotated
slip systems. Local lattice rotation is self consistently determined with the
evolving Burgers vector density distribution.

Coarse grained descriptions of defected crystalline lattices are often based on
Nye's dislocation density tensor \cite{re:nye53}, and have been
summarized in a number of excellent monographs
\cite{re:kosevich79,re:kroner81,re:mura87,re:nelson02}. The general starting
point is the introduction of a coarse graining volume that contains a large
number of defect lines threading it. The resulting dislocation density
tensor $\alpha_{ik}$ depends on the distribution of geometrically necessary
dislocations in the volume, while statistically stored dislocations (those
portions of dislocation loops that do not contribute to the dislocation density
tensor) are averaged out in the coarse-graining \cite{re:arsemlis99}. In three
dimensions, the dislocation density tensor is $\alpha_{ik} = - \epsilon_{ilm} \partial_{l} {\rm
w}_{mk}$ where $\epsilon_{ilm}$ is the anti symmetric Levi-Civita tensor, and
${\rm w}_{mk} = \partial_{m} u_{k}$ is the elastic distortion tensor. The
dislocation density tensor can be represented by a vector in two
dimensions which we refer to as the Burgers vector density
$\mathbf{b}(\mathbf{r})$. In the $\mathbf{r} = (x,y)$ plane $b_{k} (\mathbf{r})
= \alpha_{3k}(\mathbf{r})$ and hence can be written as $b_{k} =
\epsilon_{ml}\partial_{l} {\rm w}_{mk}$ where $\epsilon_{ml}$ is the
two dimensional anti symmetric tensor.

Our approach follows closely the particular description employed in equilibrium
theories of two dimensional melting
\cite{re:nelson79,re:zippelius80,re:nelson81}. In addition to the strain, the
primary variables employed to describe this two dimensional defected medium
include the Burgers vector density $\mathbf{b}(\mathbf{r})$ and the local
(coarse grained) bond angle field $\theta(\mathbf{r})$ (also called lattice
rotation). The system is assumed to be in elastic equilibrium at all times
consistent with a given defect distribution, so that strain and bond orientation
fluctuations are slaved to the instantaneous defect density distribution.
Equilibrium fluctuations in $\theta(\mathbf{r})$ were computed within linear
elasticity in Ref. \cite{re:nelson81}, and shown not to destroy long
range orientational bond order in a two dimensional crystalline lattice.

The same coarse grained description together with the methods of linear irreversible
thermodynamics have been used to obtain the equations governing dissipative
motion of the dislocation density tensor under the assumption that it is driven
by free energy minimization
\cite{re:groma97,re:rickman97,re:zaiser01,re:limkumnerd06,re:acharya10}.  We
extend this research here by incorporating a defect mobility that explicitly
depends on variations in the local orientation of the slip lines in the defected
medium.


Our study is motivated by recent developments that allow quantitative
characterization of defect structures and motion at the nanoscale. For
example, recent high resolution microscopy studies have enabled imaging of the
displacement fields created by dislocations with sub Angstrom resolution
\cite{re:hytch03}. At the same time, equilibrium configurations
\cite{re:pertsinidis01} and defect motion \cite{re:pertsinidis01b} have been
investigated in a special realization of a two dimensional crystal: a colloidal
lattice. This system affords convenient visualization of defect configurations
and the concomitant strain fields. In particular, optical tweezers methods have
recently allowed a very detailed analysis of the microscopic mechanisms of
defect motion, including the emergence of dissipative motion as the extent of
the defect increases \cite{re:irvine13}. The nanoscale structure of
isolated defects has also been recently resolved in smectic liquid
crystals with cryo-electron microscopy \cite{re:zhang15}, with some
surprising results concerning the structure and extent of edge dislocations.
Additional interest in defect motion in two dimensional systems has
been spurred by novel strain engineering methods that seek to control the electronic properties of graphene sheets
\cite{re:guinea10,re:chen11,re:bonilla12}.

Our study is also motivated by fully microscopic numerical investigations of a
variety of defect mediated dynamics, including, for example, interactions among
an ensemble of dislocations \cite{re:lesar02}, plastic deformation or grain
boundary motion \cite{re:gulluoglu90}. Simple early models of plastic
deformation in metals that are based on the existence of Frank-Read dislocation
sources and their glide over lattice-specific slip planes have been greatly
extended thanks to information obtained through massively parallel Molecular
Dynamics studies. Such atomistic level simulations have enabled quantitative
descriptions of complex situations in heavily deformed materials, e.g.,
dislocation nucleation at grain boundaries and their coupled motion
\cite{re:yamakov02}.
Although atomistic in scale, the simulations methods are largely based on
dissipative (or non-inertial) motion. This is accomplished by the introduction
of suitable \lq\lq thermostats" in the simulations, or by explicitly solving an
elastic boundary value problem slaved to the instantaneous location of the
defect lines \cite{re:crone14}. The general assumption is that defect segment
motion occurs in a time scale that is much slower than the characteristic time
of elastic relaxation of the medium. This separation of time scales is also
implicit in the model described in this paper. The model which we describe aims
at a coarse grained description of these simulations while still retaining
mesoscale information about the lattice slip planes and their contribution to
defect motion.

Bridging experiments at the nanoscale and related microscopic
numerical studies with macroscopic descriptions based
on continuum elasticity theory has proven difficult, but doing so is becoming a
necessity in order to properly describe microstructural evolution in
nanostructured materials
\cite{re:yamakov02,re:limkumnerd06,re:groger08,re:li08,re:chen13}.  We do not
attempt here a derivation of dislocation mobilities from a microscopic model of
a two-dimensional lattice. Rather we use symmetry arguments to propose
phenomenological equations of defect motion that depend on the symmetry and
local state of orientation of the lattice, and that allow for differential
defect motion along distinguished directions. We consider two possible types of
crystalline lattices in two dimensions: hexagonal and square. In the former
case, the description is somewhat simpler in that, to linear order,
defect energies are the same as in an isotropic material. However, the
description of lattice effects near defect cores is complicated by the
need to introduce geometrically unnecessary dislocations. On the
square lattice, on the other hand, the anisotropic nature of the
linear response is more complex. In both cases we obtain the
orientation dependent mobilities under several approximations. We
close by presenting an illustrative example involving the motion of
two edge dislocations. We make a number of simplifications to make the
calculation analytically tractable, and show how lattice rotation
affects glide and climb motion, and how it can prevent dislocation
annhilation thorough the local distortion of the slip planes.

\section{Mesoscopic models}
\label{sec:mesoscopic}

We consider a two dimensional crystal that contains a large number of
dislocations which are relatively close to each other, yet separated by
distances much larger than the lattice spacing so that the distribution can
be effectively coarse grained. A coarse graining cell is
introduced with a net Burgers vector that is the sum of the many Burgers vectors
of the underlying crystal dislocations within the cell. As is standard
(see, e.g., Refs. \cite{re:kosevich79,re:nelson81}), the resulting Burgers vector
density is
approximated by a continuous vector field $\mathbf{b}(\mathbf{r})$ on this two
dimensional space (with components $b_{i}(\mathbf{r}) = \alpha_{3i}(\mathbf{r}),
i=x, y$ and 3 denoting the direction perpendicular to the
plane). We first decompose the Burgers vector density into
a combination of a finite number of discrete slip systems
\cite{re:kosevich79,re:zippelius80,re:acharya01,re:groger08}
\begin{equation}
\mathbf{b}(\mathbf{r}) =
\sum_{s} b^{(s)}(\mathbf{r}) \bm{\hat{\theta}}^{(s)}(\mathbf{r})
\label{eq:b_decomp}
\end{equation}
where $s$ runs over the possible slip systems with Burgers vector density
$b^{(s)}$ locally oriented along the direction
$\bm{\hat{\theta}}^{(s)}(\mathbf{r})$. We assume that the unit vectors
$\bm{\hat{\theta}}^{(s)}(\mathbf{r})$ can be expressed as
$\bm{\hat{\theta}}^{(s)}(\mathbf{r}) = (\cos(\theta(\mathbf{r}) + \pi s/2),
\sin(\theta(\mathbf{r}) + \pi s/2 ) )$, $s =0, 1$, for a square lattice and
$\bm{\hat{\theta}}^{s}(\mathbf{r}) = (\cos(\theta(\mathbf{r}) + 2 \pi s/3),
\sin(\theta(\mathbf{r}) + 2 \pi s/3 ) )$, $s =0, 1, 2$, for a hexagonal lattice.
The local rotation of the coarse graining cell is $\theta(\mathbf{r}) = (1/2)
\epsilon_{ij} {\rm w}_{ij}(\mathbf{r})$ where $\epsilon_{ij}$ is the anti
symmetric Levi-Civita tensor, and ${\rm w}_{ij}$ the elastic distortion tensor. The
lines defined by the directions $\bm{\hat{\theta}}^{(s)}(\mathbf{r})$ do not
cross if there are no unbound disclinations \cite{re:Bowick01}, which we assume
throughout this paper.

In an unbounded medium, it is possible to express the elastic energy as a
function of the Burgers vector density. For an isotropic system, this energy is
given by \cite{re:Nelson78,re:Chaikin95} 
\begin{eqnarray}
H_{\textrm{int}} = -\frac{K}{2} \int_{|\mathbf{r}-\mathbf{r}'|>a} d \mathbf{r} d \mathbf{r}'
\left[\mathbf{b}(\mathbf{r}) \cdot \mathbf{b}(\mathbf{r}')
  \ln\left({\frac{\rho}{a}}\right) -  \mathbf{b}(\mathbf{r}) \cdot
  \bm{\hat{\rho}} \; \mathbf{b}(\mathbf{r'}) \cdot \bm{\hat{\rho}}
\right], 
\label{Hamiltonian}
\end{eqnarray}
where $a$ is a short distance cutoff on the order of the lattice spacing,
$K$ is the two dimensional Young's modulus and $\bm{\rho} = \mathbf{r} - \mathbf{r}'$,
$\bm{\hat{\rho}}$ the corresponding unit vector, and $\rho = \| \bm{\rho} \|$.
This expression does not include a nonlocal self-energy of the dislocation
distribution due to their long ranged strain field because the total Burgers
vector over the entire system is taken to be zero, so that dislocations are
created and annihilated in opposing pairs. There is also, however, a local
energy contribution associated with the nonlinear strain fields near the core of
the dislocation. This energy is assumed to be approximately independent of the
local strain field due to other sources \cite{re:Fisher79}, and is modeled
by a quadratic term in the Burgers vector \cite{re:Nelson78} 
\begin{equation}
\label{eq:loc}
H_{\textrm{loc}} = E_{c} \int d \mathbf{r} \; \mathbf{b}(\mathbf{r})
\cdot \mathbf{b}(\mathbf{r}),
\end{equation}
with $E_{c}$ a constant core energy. Below we will propose a slightly different
core energy to also include the energy of geometrically unnecessary dislocations
(dislocation groups that do not contribute to the local Burgers vector density).

In an unbounded system, the solution of the equilibrium elasticity problem is
equivalent to obtaining the Burgers vector density distribution. This is because
the incompatibility of the plastic strain is completely balanced by an elastic
strain that makes the total strain compatible \cite{re:kroner81}. This allows
one to express the solution for the strain field as a function only of the
Burgers vector density that acts as a source of strain
\cite{re:mura87,re:Chaikin95}. 

Dislocations and other defects play a key role in determining the evolution,
properties, and response of materials outside of thermodynamic equilibrium.
While the systems under study here are assumed to be in elastic equilibrium
relative to a given defect distribution, defects interact, and are free to move
and annihilate to relieve stresses and reduce the overall energy of the system.
Such an evolution can have reversible and irreversible contributions that
correspond to different models of relaxation \cite{re:irvine13}. 
A number of theoretical studies in the literature have addressed dissipative
motion of an ensemble of dislocations at the mesoscale
\cite{re:groma97,re:rickman97,re:aguenaou97,re:zaiser01,re:haataja02,re:limkumnerd06,re:limkumnerd07}. 
A relaxational equation for the Burgers vector density is introduced under the
assumption that the evolution of the density is driven by plastic energy
minimization. The equation is of the general form,
\begin{eqnarray}
\frac{\partial b_j}{\partial t} = -\epsilon_{lm} B_{mjsi}
\epsilon_{sb} \partial_l \partial_b \frac{\delta H}{\delta b_{i}}, 
\label{eq:burger_evolution}
\end{eqnarray}
where $H = H_{int} + H_{loc}$, and $B_{mjst}$ is a constant mobility tensor. We
propose in this paper a more accurate description of the kinetic motion of the
defect distribution by considering anisotropic mobilities along slip
lines of the lattice
rather than along the orientations defined locally by the Burgers vector
density as is the case in Eq. (\ref{eq:burger_evolution}). Moreover, we show how to
distinguish glide and climb in two dimensional lattices that are locally
rotated, as they are in the presence of an ensemble of dislocations.

Within linear elasticity in an isotropic medium the local orientation of a
two dimensional coarse graining cell is related to the Burgers vector density
through a nonlocal relation \cite{re:dewit73}, 
\begin{eqnarray}
\theta(\mathbf{r}) = -\frac{1}{2 \pi} \int d \mathbf{r}' \;
\frac{\mathbf{b}(\mathbf{r}') \cdot \bm{\hat{\rho}}}{\rho}. 
\label{theta}
\end{eqnarray} 
On an infinite lattice in which the Burgers vector decays sufficiently fast at
infinity we can take $\theta(\mathbf{r}) = 0$ at infinity \cite{fo:bp1_01}. The
fact that the orientation is different at all points on the plane implies that
the local slip lines $\bm{\hat{\theta}}^{(s)}(\mathbf{r})$ are also position
dependent.  Therefore if the dislocation mobility is anisotropic, Eq.
(\ref{eq:burger_evolution}) will not adequately describe defect motion along
locally rotated slip systems. 


We propose to extend Eq. (\ref{eq:burger_evolution}) in two ways, both
phenomenological and based on symmetry arguments. First, in the presence of an
orientation field $\theta(\mathbf{r})$, or lattice torsion, there is no longer
strict translational symmetry, but the composition of a translation and a
rotation due to plastic deformation. In this way, the configurational energy
depends explicitly on local orientation, as the lattice symmetries of reflection
and rotation must be applied locally \cite{re:nelson81}.  Second individual
dislocations respond anisotropically to forcing so that the motion of an
ensemble of dislocations depends on how the local Burgers vector
density is
decomposed among slip systems as shown in Eq. (\ref{eq:b_decomp}). We note that while we allow the slip system
directions to be different from one coarse graining cell to another, we 
neglect changes to the relative angle between them due to deformation of
the cell. Hence the local coordinate axis system defined by the slip systems
$\bm{\hat{\theta}}^{(s)}(\mathbf{r})$ is, approximately, determined by a single
angle $\theta(\mathbf{r})$ (as explicitly shown below Eq. (\ref{eq:b_decomp})).



\section{Dislocation motion on a square lattice}
\label{sec:square}

The symmetry of the square lattice is generated by rotations about $\pi/2$ and
reflections about the two bond axes forming the group $D_4$. This symmetry
implies that a rank two tensor (a matrix) relating two vectors transforming
under $\textrm{SO}(2)$ has to be proportional to the identity matrix. This can
be checked by assuming the most general $2\times 2$ matrix and applying the
transformation matrices, demanding equality of the initial and transformed
matrices. A similar analysis for the compliance matrix, a rank four tensor
relating the stress matrix to the strain matrix within linear elasticity
(Hooke's Law) $u_{ij} = S_{ijkl} \sigma_{kl}$, shows that it can be written in
general as,
\begin{eqnarray}
S_{ijkl} = \alpha \delta_{ij}\delta_{kl} + \beta
\delta_{i(k}\delta_{l)j} + \Delta \delta_{ijkl},
\label{eq:tensor_decomposition}
\end{eqnarray}
where $\alpha, \beta$, and $\Delta$ are constants related to the elastic
constants of the lattice, and $\delta_{ijkl}$ is the fourth rank identity
tensor. Here and below we will make use of the notation $A_{(bc)} =
\frac{1}{2}(A_{bc} + A_{cb})$ and $A_{[bc]} = \frac{1}{2}(A_{bc} - A_{cb})$.
There is an additional term allowed for a general fourth rank tensor which is
not present here because the stress is symmetric $\sigma_{ij}=\sigma_{ji}$. 

In the case of hexagonal symmetry addressed in
Sec. \ref{sec:hexagonal}, invariance under rotations of $\pi/3$ and
reflections about the three independent bond orientations, and again within
linear elasticity, leads to the same decomposition
(\ref{eq:tensor_decomposition}),
but with $\Delta = 0$. Note that in this case, and within linear
distortions, the compliance matrix has the same decomposition as in an isotropic
system. In this latter case, the system is invariant under arbitrary rotations
and reflections.

An approximate expression for the energy of a distorted and rotated lattice
can be obtained by applying the tensorial decomposition above in the local
coordinates of each rotated coarse graining cell (it is still the case that the
stress is symmetric on its indices in these coordinates). Introduce a local
coordinate system with unit vectors $\mathbf{\hat{x}}'$ and $\mathbf{\hat{y}}'$
that are related to laboratory coordinates $x$ and $y$ by a rotation about
$\theta(\mathbf{r})$: $\mathbf{\hat{x}}' = (\cos \theta , \sin \theta )^T$ and
$\mathbf{\hat{y}}' = (- \sin \theta , \cos \theta )^T$. We use in what follows
upper indices for tensors expressed in local coordinates and lower indices for
tensors in the laboratory frame. Then, for example, $\sigma^{ab} =
R^{a}_{.i}[\theta(\mathbf{r})] R^{b}_{.j}[\theta(\mathbf{r})] \sigma_{ij}$,
where we have introduced the rotation matrix
\begin{equation}
R^{a}_{.i} [\theta] = \left(\begin{array}{cc}
\cos \theta & -\sin \theta \\
\sin \theta & \cos \theta
\end{array}\right).
\label{eq:rotation_matrix}
\end{equation}
By reason of symmetry, we have
\begin{equation} 
\label{eq:localform}
u^{ij} = \alpha \delta^{ij}\sigma^{kk} +  \beta \sigma^{ij} +
\Delta (\delta^{i x'} \delta^{j x'} \sigma^{x'x'} + \delta^{i
y'}\delta^{j y'} \sigma^{y'y'})
\end{equation}
in local coordinates. Equation (\ref{eq:localform}) transformed to the 
laboratory frame reads,
\begin{equation}
u_{ij} = \alpha \delta_{ij}\sigma_{kk} + \beta \sigma_{ij} + 
\Delta h_{ijkl}(\theta(\mathbf{r})) \sigma_{kl}.
\label{eq:strain_stress_cubic}
\end{equation}
with
\begin{eqnarray}
h_{xxxx} & = & h_{yyyy} = \cos^4 \theta + \sin^4 \theta, \quad \quad 
h_{xxyy} = \frac{1}{2}\sin^2 2\theta, \nonumber \\
h_{xxxy} & = & -h_{xyyy} = \frac{1}{4}\sin 4 \theta,  
\label{h}
\end{eqnarray}
where the other components of the tensor function $h_{klmn}$ come from that fact
that it does not depend on the order of its indices (a general result for this
symmetry). We also have used the notation $\sigma_{kk} = {\rm Tr}(\sigma_{ij})$.


The elastic energy can now be calculated as follows:
Since $\partial_j
\sigma_{ij} = 0$ and $\sigma_{ij}=\sigma_{ji}$ an Airy stress function
$\chi(\mathbf{r})$ is introduced such that 
\begin{equation} \sigma_{ij} =
\epsilon_{ik} \epsilon_{jl} \partial_k \partial_l \chi. 
\label{eq:Airy}
\end{equation}
When there are no free disclinations, it is possible to express the
Airy stress function in terms of Burgers
vector density \cite{re:Chaikin95}. Apply
$\epsilon_{ik}\epsilon_{jl}\partial_{k}\partial_{l}$ to Eq.
(\ref{eq:strain_stress_cubic}) and substitute the definition (\ref{eq:Airy}) to
find,
\begin{equation}
\epsilon_{ik} \epsilon_{jl} \partial_k \partial_l u_{ij} = \alpha' 
\nabla^4 \chi + \Delta \epsilon_{ik} \epsilon_{jl} \partial_k 
\partial_l \hat{\mathcal{D}}_{ij}\chi,
\label{eq:b_square}
\end{equation}
where we have introduced $\alpha' = \alpha + \beta$, and the differential
operator 
\begin{equation}
\hat{\mathcal{D}}_{ij}[\theta] = h_{ijkl}(\theta) \epsilon_{km}
\epsilon_{ln} \partial_m \partial_n.
\label{eq:Dij}
\end{equation}
The left hand side of Eq. (\ref{eq:b_square}) is, by definition,
$\epsilon_{ik} \epsilon_{jl} \partial_k \partial_l u_{ij} =
\epsilon_{ij}\partial_i b_j$.  This definition, together with Eqs.
(\ref{eq:b_square}) and (\ref{eq:Dij}), is the solution of the equilibrium
elastic problem that gives $\chi(\mathbf{r})$ as a function distribution of the
Burgers vector density $\mathbf{b}(\mathbf{r})$ and rotation
$\theta(\mathbf{r})$ that still remains to be determined.


Once the solution $\chi(\mathbf{r})$ is determined, the energy of the
configuration $H_{\textrm{int}} = \frac{1}{2}\int d^2 r u_{ij} \sigma_{ij}$ can
be found by substituting Eq. (\ref{eq:strain_stress_cubic}) for the strain, and
the definition of the Airy function, Eq. (\ref{eq:Airy}), for the stress. We
find,
\begin{equation}
H_{\textrm{int}} = \frac{1}{2}\int d^2 r \chi(r) \left[ \alpha' \nabla^4 + 
\Delta \epsilon_{ik} \epsilon_{jl} \partial_k \partial_l 
\hat{\mathcal{D}}_{ij} \right] \chi(r),
\label{eq:energy_square}
\end{equation}

Equation (\ref{eq:b_square}) cannot be solved explicitly for the Airy
function, and hence we cannot express the energy
(\ref{eq:energy_square}) explicitly as a function of the
Burgers vector density, unlike the isotropic case of $\Delta = 0$ (in this
latter case, the differential equation (\ref{eq:b_square}) is solved by using a
Green's function method, see Nelson in his seminal paper \cite{re:Nelson78},
leading to Eq. (\ref{Hamiltonian}) for the energy of interaction). Furthermore,
the energy depends on the rotation $\theta$ through the dependence of the
differential operator $\hat{\cal D}_{ij}$, Eq. (\ref{eq:Dij}). Obtaining such a
relation is the subject of the next subsection.

Before proceeding, we note that it is possible to find a closed form
of the energy if rotation is neglected, and one starts from the
general form of Hooke's law for a square lattice, Eq.
(\ref{eq:localform}), written in laboratory frame coordinates
(the linear elasticity regime, see, e.g., \cite{re:steeds73}). Since 
\begin{equation}
\hat{\cal D}_{ij} (\theta = 0) =
\left[\delta_{ix}\delta_{jx} \epsilon_{xl} \epsilon_{xk}  + 
\delta_{iy}\delta_{jy} \epsilon_{yl} \epsilon_{yk}\right] 
\partial_l \partial_k \chi = \left[\delta_{ix}\delta_{jx} \partial_y^2  + 
\delta_{iy}\delta_{jy} \partial_{x}^{2}\right] \chi , 
\end{equation}
Eq. (\ref{eq:energy_square}) reduces to
\begin{equation}
H_{\textrm{int}} = \frac{1}{2}\int d^2 r \chi(r) \left[ (\alpha'+ \Delta) 
\nabla^4 - 2 \Delta \partial_x^2 \partial_y^2 \right] \chi(r).
\label{eq:energy_ordered}
\end{equation}
After Fourier transformation, substitution of Eq. (\ref{eq:b_square})
into Eq. (\ref{eq:energy_ordered}) leads to an explicitly form of the
energy in terms of the Burgers vector density
\begin{equation}
H_{\textrm{int}} = \frac{1}{2}\int \frac{d^2 q}{(2 \pi)^2} \frac{|i 
\epsilon_{ij} q_i b_j|^2}{(\alpha'+\Delta) q^4 - 2 \Delta q_x^2 q_y^2}
= \frac{1}{2}\int \frac{d^2 q}{(2 \pi)^2} \frac{\left( q^2 \delta_{ij} - 
q_i q_j\right) }{(\alpha'+\Delta) q^4 - 2 \Delta q_x^2 q_y^2} b_i(q) b_j(-q).
\label{eq:hamiltonian_ordered}
\end{equation}
This extends the isotropic result of $\Delta = 0$ to the
square lattice.


\subsection{Lattice rotation field}

We next determine the nonlocal relationship between the local rotation of a
coarse graining cell and the Burgers vector distribution to generalize
Eq. (\ref{theta}) to a square lattice. The local rotation
$\theta(\mathbf{r})$, relative to an
undistorted reference lattice with $\theta =0$, is related the distortion tensor
${\rm w}_{ij}$. The symmetric and anti symmetric parts of the distortion tensor
are identified as the strain and orientation tensors respectively
\cite{re:kroner81}
\begin{equation}
{\rm w}_{ij} = u_{ij} + \theta(\mathbf{r}) \epsilon_{ij}.
\label{eq:w_decomp}
\end{equation}
By recalling the definition of the Burgers vector density in terms of the
distortion tensor $b_{k} = \epsilon_{ij}\dd_j {\rm w}_{ik}$, 
and substituting the decomposition of the distortion tensor, Eq.
(\ref{eq:w_decomp}), one has 
\begin{equation} 
\label{eq:dist}
b_k = \epsilon_{ij}\partial_j \left( \theta \epsilon_{ik} + u_{ik} \right) 
= \partial_k \theta + \epsilon_{ij} \partial_j u_{ik}.
\end{equation}
Thus up to a constant, $\theta$ is specified by $\partial_k \theta = b_k -
\epsilon_{ij} \partial_j u_{ik}$. 

The divergence of second term in the r.h.s. of Eq. (\ref{eq:dist}) can
be calculated with the help of Eqs. (\ref{eq:Airy}) and (\ref{eq:Dij}),
\begin{equation}
\partial_k \epsilon_{ij} \partial_j u_{ik} = \alpha
\left(\epsilon_{ij} \partial_j \partial_i\right) \sigma_{ll} +
\beta \epsilon_{ij} \partial_j \left(\partial_k \sigma_{ik}\right)
+ \Delta \epsilon_{ij} \partial_k \partial_j \hat{\mathcal{D}}_{ik}
\chi =\Delta \epsilon_{ij} \partial_k \partial_j  \hat{\mathcal{D}}_{ik} \chi,
\end{equation}
where we have used the anti symmetry of $\epsilon_{ij}$ and the condition of
elastic equilibrium $\partial_k \sigma_{ik} = 0$. Thus the divergence of Eq.
(\ref{eq:dist}) is given by,
\begin{eqnarray*}
\nabla^2 \theta = \partial_k b_k + \Delta \epsilon_{ij} \partial_k 
\partial_j \hat{\mathcal{D}}_{ik} \chi.
\end{eqnarray*}
To solve for $\theta$ we introduce the Green's function of the two dimensional
Laplacian operator and find,
\begin{equation}
\theta(\mathbf{r}) = \frac{1}{2 \pi} \int_{|\mathbf{r}-\mathbf{r}'|>a} 
d \bm{r}^{\prime} \ln\left(\frac{|\mathbf{r}-\mathbf{r}'|}{a}\right) 
\partial_k' \mathcal{B}_k (\theta,\mathbf{r}') 
= -\frac{1}{2 \pi} \int_{|\mathbf{r}-\mathbf{r}'|>a} d \bm{r}^{\prime} 
\frac{r_k - r_k'}{|\mathbf{r}-\mathbf{r}'|^{2}} 
\mathcal{B}_k [\theta,\mathbf{r}'],  
\label{eq:theta_square}
\end{equation}
where
\begin{equation}
\mathcal{B}_k [\theta,\mathbf{r}'] = b_k(\mathbf{r}') + \Delta
\epsilon_{ij} \partial_j'
\hat{\mathcal{D}}_{ik}(\theta(\mathbf{r}')) \chi(\mathbf{r}'),
\label{eq:calB}
\end{equation}
which reduces to the Burgers vector density of an isotropic system when
$\Delta=0$. 

Equations (\ref{eq:b_square}), (\ref{eq:theta_square}), and (\ref{eq:calB}), now
constitute a closed set of equations for the elastostatics of a square lattice
in terms of $\theta$ and $\mathbf{b}$. Equation (\ref{eq:theta_square}),
however, is only an implicit equation for $\theta(\mathbf{r})$. As pointed out
by Kr\"oner \cite{re:kroner81}, to obtain a relation between $\mathbf{b}$ and
$\theta$ one must solve the problem of elastic equilibrium everywhere in order
to relate the stress $\sigma_{ij}$ to the Burger's vector density $\mathbf{b}$. 

A simpler form follows if $\theta$ is
everywhere small so that it can be approximated by a constant in the right-hand
side of Eq. (\ref{eq:theta_square}). Then
\begin{equation}
\label{theta_ordered}
\theta(\mathbf{r}) = -\frac{1}{2 \pi}
\int_{|\mathbf{r}-\mathbf{r}'|>a} \frac{d
  \bm{r}^{\prime}}{|\mathbf{r}-\mathbf{r}'|^{2}} \left[ (r_k - r_k')
  b_k(\mathbf{r}') +  \Delta (x-x')(y-y') \left(
    \sigma_{xx}(\mathbf{r}') - \sigma_{yy}(\mathbf{r}')\right)\right],
\end{equation}
where $a$ is a short distance cutoff on the order of the lattice spacing and we
have dropped some boundary terms. This reduces to Eq.  (\ref{theta}) in the
isotropic limit $\Delta = 0$.

\subsection{Dynamics}

We extend next the kinetic equation (\ref{eq:burger_evolution}). We decompose
the Burgers vector density into a finite number of slip systems
$\mathbf{b}(\mathbf{r}) = \sum_{s} b^{(s)}(\mathbf{r})
\bm{\hat{\theta}}^{(s)}(\mathbf{r})$, each defined by its own density $b^{(s)}$
oriented along the direction $\bm{\hat{\theta}}^{(s)}(\mathbf{r})$.  On the
square lattice we simply have $b^{(s)} = b^{s}$ and $\bm{\hat{\theta}}^{(s)} =
\bm{\hat{\theta}}^{s}$, the variables along the locally rotated coordinate
system.

Since the Burger's vector is a pseudo-vector (it is even under a parity
transformation, whereas a vector is odd under parity) the natural Burgers vector
flux is a pseudo-tensor $\Phi_k^i$, which represents the flux along the
$k$-direction of dislocations along the $i$-direction. For simplicity, we limit
our analysis here to the case in which the Burgers vector
densities are separately conserved \cite{re:haataja02}
\begin{eqnarray}
\frac{\partial b^{i}(\mathbf{r})}{\partial t} = - \partial_k \Phi_{k}^{i}.
\label{cons_sq}
\end{eqnarray}
Explicitly, the assumption is that dislocations can only be created or destroyed
by pair annihilation and creation on each slip system. This requirement also
guarantees that the energy integral is finite for an infinite system. As stated
earlier, we require that the bond directions are well defined, which implies the
absence of free disclinations \cite{re:Bowick01}. 

Thermodynamic forces leading to defect motion arise from 
$\frac{\delta H}{\delta b^{i}(\mathbf{r})}$, the change in energy for a
dislocation along slip plane direction $i$ to be placed at $\mathbf{r}$.
Therefore its partial derivative $\partial_k \frac{\delta H}{\delta
b^{i}(\mathbf{r})}$ represents the local difference in energy for dislocation
placement, and is thus the thermodynamic force. The total energy $H$ is
quadratic in $b^{i}$ so that the resulting thermodynamic force will be linear in
$b^{i}$, although nonlocal. Since $b^{i}$ is pseudo-scalar, we find that the
thermodynamic force is a pseudo-scalar. 
In linear response, forces and fluxes are linearly related as, 
\begin{eqnarray}
\Phi_{k}^{i}(\mathbf{r}) = - D^{i}_{kj}(\mathbf{r}) \;\; \dd_j
\frac{\delta H}{\delta b^{i}(\mathbf{r})}.
\label{linear_square}
\end{eqnarray}
This expression is nonlocal because the thermodynamic force is a nonlocal
functional of the dislocation densities. Of course, this is only the case in the
slow temporal scale of dislocation segment motion, and is a consequence of the
assumption that the system is at all times in elastic equilibrium.
We now distinguish glide and climb motion and decompose $D_{kj}^{i}$ along the
direction $\bm{\hat{\theta}}^{i}(\mathbf{r})$ and transverse to it
\begin{equation}
D^{i}_{kj} = D_{g}\theta^{i}_{k} \theta^{i}_{j} + D_{c}
\left(\delta_{kj}-\theta^{i}_{k} \theta^{i}_{j}\right), 
\label{coefficients}
\end{equation}
where $D_g$ is identified as the mobility for glide motion, and $D_c$ for climb.
For a square lattice, we write
\begin{eqnarray}
\bm{\hat{\theta}}^{l}(\mathbf{r}) = \left(\begin{array}{c}
\cos[\theta(\mathbf{r}) + l\frac{\pi}{2}] \\
\sin[\theta(\mathbf{r})+ l\frac{\pi}{2}] 
\end{array}\right), 
\label{orientation_sq}
\end{eqnarray}
where $l = 0$ defines $x'$ and $l=1$, $y'$. A similar decomposition
of the dislocation mobility into climb and glide components was given in the
study of elastic instabilities of thin films \cite{re:haataja02}, and for the
motion of isolated dislocations \cite{re:zippelius80,re:tsimring95}.

By combining Eqs. (\ref{cons_sq}), (\ref{linear_square}), and
(\ref{coefficients}) we obtain the phenomenological equation of motion for the
Burgers vector densities,
\begin{equation}
\frac{\partial b^{i}(\mathbf{r})}{\partial t} = \left[ \partial_{k}
  (D_{g} - D_{c}) \theta_{k}^{i} \theta_{j}^{i} \partial_{j} 
 + D_c \nabla^2 \right] \frac{\delta H}{\delta b^{i}(\mathbf{r})}.
\label{EOM_square}
\end{equation}
This dynamical equation along with Eqs. (\ref{eq:b_square}),
(\ref{eq:energy_square}), (\ref{eq:theta_square}), and (\ref{eq:calB})
completely specify our anisotropic model on the square lattice. This, and the
corresponding expression for a hexagonal lattice to be given below, are the
central results of this paper.

Prior work has not considered lattice rotation effects on dislocation
motion. We briefly show that Eq. (\ref{EOM_square}) reduces to simpler expressions,
already in the literature, when rotation is uniform.
This simpler description allows for a more direct comparison with
isotropic theories in which the laboratory coordinate system is the
natural choice. 
We begin by writing
\begin{eqnarray}
\partial_t b_k = \partial_t \sum_i \hat{\theta}^{i}_{k} b^{i} 
= \sum_{i} \hat{\theta}^{i}_{k} \partial_t b^{i},
\end{eqnarray}
Then inserting Eq. (\ref{EOM_square}), we find
\begin{eqnarray}
\partial_t b_k =  \sum_i \hat{\theta}^{i}_k \partial_n
D^i_{nm}  \partial_m \frac{\delta H}{\delta b^{i}}. 
\end{eqnarray}
We also have the relation
$\frac{\delta H}{\delta b^{i}} = \hat{\theta}^{i}_l \frac{\delta H}{\delta
b_l}$, which follows from the chain rule. Then we can write the response
explicitly in terms of the Burgers vector density alone,
\begin{equation}
\partial_t b_k  = \sum_i \hat{\theta}^{i}_k D^i_{nm}
\hat{\theta}^{i}_l \left\{ \partial_n \partial_m   \frac{\delta H}{\delta b_l} \right\},
\label{eq:burgers_kinetic}
\end{equation}
which explicitly separates the current originating from the excess energy
associated with dislocations and a mobility coefficient that depends on local
orientation. Substitute Eq. (\ref{coefficients}) into Eq.
(\ref{eq:burgers_kinetic}) and evaluate the sums over the orientation directions
\begin{eqnarray}
\sum_{i} \theta^{i}_k \theta^{i}_l = \delta_{kl}, \quad \quad
\sum_{i} \theta^{i}_k \theta^{i}_l \theta^{i}_m \theta^{i}_n = h_{klmn}(\theta),
\end{eqnarray}
where the rank four tensor $h$ is defined in Eq. (\ref{h}),
so that Eq. (\ref{eq:burgers_kinetic}) reduces to
\begin{equation}
\partial_t b_k = K \left[ (D_g-D_c) h_{kmnl} + D_c
  \delta_{mn}\delta_{kl} \right] \partial_n \partial_m  \frac{\delta
  H}{\delta b_l}. 
\label{eq:final_square}
\end{equation}
Just taking the rotation angle to be zero, the equations of motion
reduce to \cite{re:steeds73}
\begin{eqnarray*}
\partial_t b_{x}(\mathbf{q},t) & = & - \left[D_g q_x^2 + D_c q_y^2\right] 
\frac{ q_y \left[q_y b_x(\mathbf{q}) - q_x b_y(\mathbf{q})\right]
}{(\alpha'+\Delta) q^4 - 2 \Delta q_x^2 q_y^2}  \nonumber \\
\partial_t b_{y}(\mathbf{q},t) & = & - \left[D_g q_y^2 + D_c q_x^2\right] 
\frac{ -q_x \left[q_y b_x(\mathbf{q}) - q_x b_y(\mathbf{q})\right]
}{(\alpha'+\Delta) q^4 - 2 \Delta q_x^2 q_y^2}.
\end{eqnarray*}

\section{Dislocation motion on a hexagonal lattice}
\label{sec:hexagonal}

The fact that the linear elastic response of a hexagonal lattice is isotropic
makes the evaluation of the elastic energy in
Eq. (\ref{eq:energy_square}) much simpler because any
dependence on lattice orientation vanishes at this order in the strain.
On the other hand, on a two dimensional hexagonal lattice there are
three independent slip planes along which individual Burgers vectors
can be oriented. As was the case for the
square lattice, the Burgers vector distribution can be written as
$\mathbf{b}(\mathbf{r}) = \sum_{s} b^{(s)}(\mathbf{r})
\bm{\hat{\theta}}^{(s)}(\mathbf{r})$ with \cite{re:zippelius80}
\begin{equation}
\bm{\hat{\theta}}^{(s)}(\mathbf{r}) = \left(\begin{array}{c}
\cos[\frac{2\pi s}{3} + \theta(\mathbf{r})] \\
\sin[\frac{2\pi s}{3} + \theta(\mathbf{r})]
\end{array}\right) \hspace{.7 cm} s = 0,1,2.
\label{orientation}
\end{equation}


Unlike the case of a square lattice, a two dimensional hexagonal lattice has
three separate slip systems, and hence three separate Burgers vector densities.
This implies that the two dimensional Burgers vector density has to be
decomposed along three independent projections, not two.
\begin{figure}
\centering
\includegraphics[width=8.6cm]{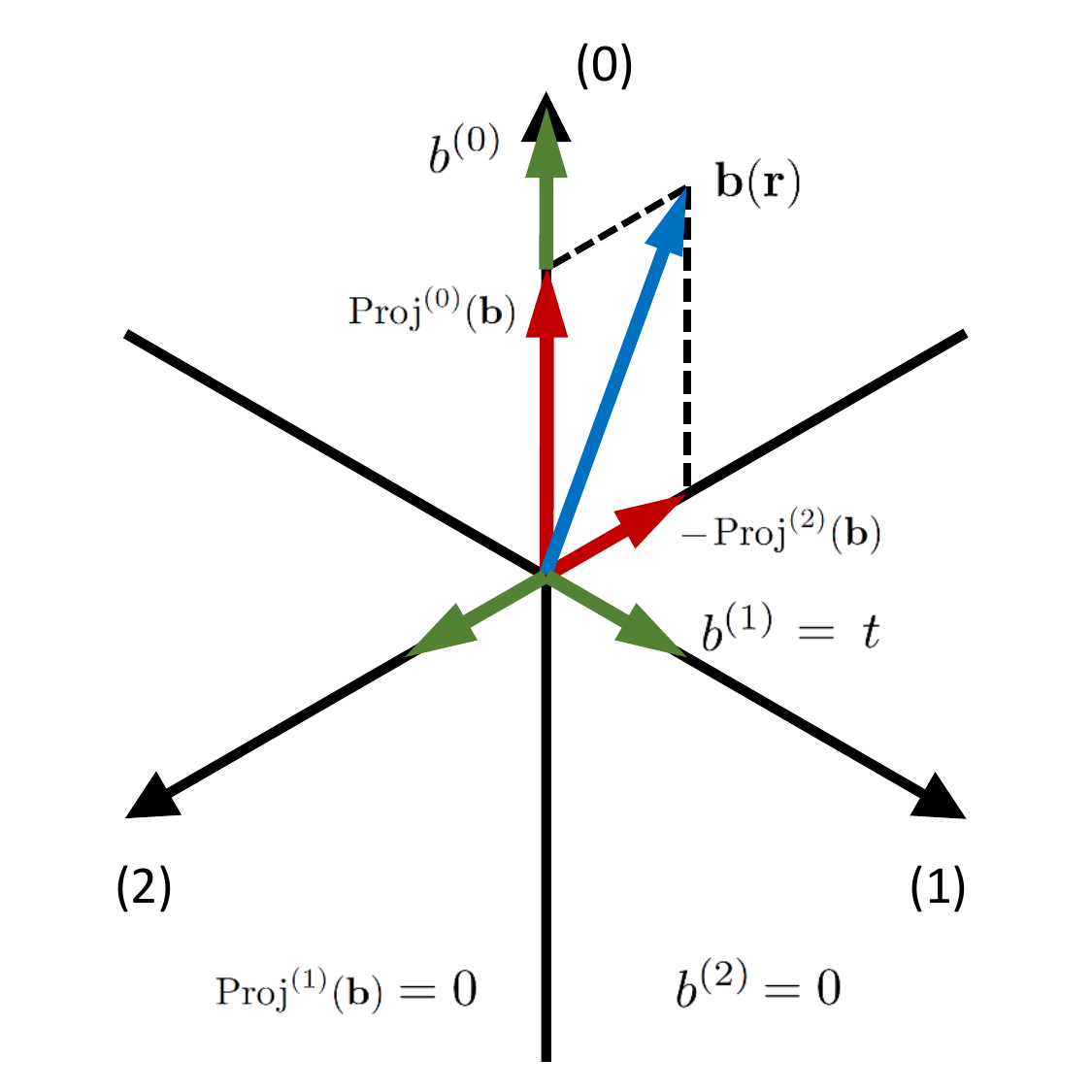}
\caption{An illustration of the two decompositions used (color
  online). The Burger's vector is shown in blue and its projections
  onto the nearest glide planes are in red. The green vectors
  represent the triplet density. The dislocation densities are the sum
  of the corresponding projection (red) and the triplet density
  (green). The case shown is when the dislocation density $b^{(2)}$
  vanishes and $b^{(1)}$ does not receive a
  projection. $\hat{\theta}^{(0)}$ is shown vertical for convenience,
  but its orientation with respect to $\hat{x}$ is given by $\theta$.} 
\label{fig:tdd}
\end{figure}
To solve this difficulty, we propose to introduce a new coarse grained field
that captures dislocation configurations not describable by the Burgers vector
density. For instance, a dislocation triplet within a single coarse graining
cell, one in each of the positive $\bm{\hat{\theta}}^{(s)}(\mathbf{r})$
directions, has zero Burgers vector. These dislocations are considered
geometrically unnecessary since they do not contribute to the elastic energy,
but they can be considered to contribute to the local anisotropic response. 
We therefore define the triplet density field $t(\mathbf{r})$ as
\begin{eqnarray}
b^{(s)} = t + \textrm{Proj}^{(s)}(\mathbf{b}),
\label{triplet}
\end{eqnarray}
with 
\begin{equation}
\textrm{Proj}^{(s)}(\mathbf{b}) = \left| \hat{\theta}^{(s)}_i
  \left(\delta_{ij} - \frac{1}{\sqrt{3}} \epsilon_{ij} \right) b_j
\right| \times \textrm{sgn} \; \left(\bm{\hat{\theta}}^{(s)}\cdot
  \mathbf{b}\right)  \Theta\left[\| \bm{\hat{\theta}}^{(s)}\cdot
  \mathbf{b} \| -\frac{1}{2}\right], 
\end{equation}
being the local projection of the Burgers vector density onto the {\em
nearest} two directions $\bm{\hat{\theta}}^{(s)}$. This term can almost be
written as a tensor; the sign function arises from the fact that our choice of
non orthogonal axes depends on the angle of $\mathbf{b}$, and the step function
$\Theta$ ensures that one axis does not receive a projection from $\mathbf{b}$.
This can also be written directly in terms of the absolute angle
$\omega^{(\alpha)}$ between $\bm{\hat{\theta}}^{(s)}$ and $\mathbf{b}$,
\begin{equation}
\textrm{Proj}^{(s)}(\mathbf{b}) = \| \mathbf{b} \|
\left|\frac{2}{\sqrt{3}} \cos\left(\omega^{(s)} -
\frac{\pi}{6}\right)\right| \times \textrm{sgn}(\cos \omega^{(s)})
\Theta\left[|\cos(\omega^{(s)})|-\frac{1}{2}\right].
\end{equation} 
A positive triplet has an equal Burgers vector in each of the positive
$\hat{\theta}^{(s)}$ directions. Note that $t$ is odd under rotations about
$\pi/3$, and hence also under reflections.

While it is simple to write down the Burgers vector density given the
Burgers density components along the slip systems, the inverse problem requires the
determination of the geometrically unnecessary density $t$. For simplicity, our
assumption here is that all of the geometrically unnecessary dislocation content
is in $t$. Therefore, the decomposition of the coarse-grained $\mathbf{b}$ onto
the two nearest lattice directions $\bm{\hat{\theta}}^{(s)}$ is minimal in the
following sense: If $\mathbf{b}$ is parallel or anti parallel to one of the slip
planes $\bm{\hat{\theta}}^{(s)}$, the lowest energy configuration is assumed to
be the one that only has dislocations pointing along this axis: $b^{(s)}=\|
\mathbf{b} \|$ with $b^{(r\neq s)}=0$. Otherwise, we project $\mathbf{b}$ onto
the two (of six) closest non-orthogonal directions along which a Burgers vector
can point. Then the remaining dislocation densities must form a zero-vector
configuration. This choice of decomposition is motivated because it is 
the one that
minimizes a local defect energy associated with a core energy that depends on
the number of dislocations rather than the magnitude of the Burgers vector. 

We consider only three separate signed densities, and opposite pairs within the
same coarse graining cell are assumed to annihilate; the only remaining
coarse-grained configurations of geometrically unnecessary defects are
dislocation triplets. We could, on the other hand, consider six separate
dislocation densities on the three slip systems. This may be a more accurate
description for large coarse graining cells. Here we may consider smaller coarse
graining cells so that opposite dislocations in the same cell would be unstable
to annihilation energetically. However, this ignores the fact that opposite
dislocations on different glide planes can form a stable dislocation pattern
that has no vector component and corresponds to a vacancy defect. We do not
consider this complication here and only add more variables as are necessary to
introduce a dependence of the response on the local bond orientation.

In summary, we choose the three densities $b^{(s)}(\mathbf{r})$ to be our
primitive variables. The lattice orientation field
$\theta(\mathbf{r})$ and the Burgers vector density
$\mathbf{b}(\mathbf{r})$ can be obtained by simultaneously solving
Eqs. (\ref{eq:b_decomp}), (\ref{theta}), and (\ref{orientation}). From
them the triplet density $t(\mathbf{r})$ can be obtained. We note that
the three Burgers vector densities and the
triplet dislocation density at every point in space contain all of the
information of the defected lattice configuration that we are considering.

The equilibrium linear elasticity of the hexagonal system is simple:
decompositions of rank two and rank four tensors are the same as for isotropic
systems. We can therefore use known results for isotropic systems:
Given the two dimensional Young modulus
$K$, the elastic energy of a configuration of defects is well
described by the long ranged transverse interaction of Eq. (\ref{Hamiltonian})
\cite{re:Chaikin95}.  The energy depends only on the Burgers vector density
$\mathbf{r}$, and not on the make-up of that defect density in terms of
dislocations densities $b^{(s)}(\mathbf{r})$ in the three different slip
directions. 

In terms of the core energy, we proceed by analogy with
Eq. (\ref{eq:loc}), and introduce a local
contribution to the energy from defect cores of the form
\begin{equation}
H_{loc} = \frac{E_c}{2} \int d \mathbf{r} \sum_{s} b^{(s)}(\mathbf{r})b^{(s)}(\mathbf{r}).
\label{eq:core_alpha}
\end{equation}
The energy $E_c$ is the approximate local energy cost due to lattice distortion
to have a dislocation pair along a given direction. Note that this expression
allows both geometrically necessary and unnecessary dislocations. The above
form is different than Eq. (\ref{eq:loc}), although the two definitions
coincide on the square lattice. Equation (\ref{eq:core_alpha}) predicts that
the core energy of a dislocation triplet is $\frac{3}{2}E_c$. The ratio of
$3/2$ between triplets and pairs is in good agreement with experiments on
two-dimensional colloids \cite{re:pertsinidis01} where the energies of
dislocation pairs and triplets were measured for various configurations, each
of which create a vacancy defect.

It is our expectation, however, that as the coarse graining cell size increases, the core
energy Eq. (\ref{eq:core_alpha}) would simply reduce to the core energy in Eq.
(\ref{eq:loc}). In fact, the local driving force arising from the core energy
satisfies the relation
\begin{eqnarray}
\sum_{s} \bm{\hat{\theta}}^{(s)}(\mathbf{r}) \frac{\delta
  H_{loc}}{\delta b^{(s)}(\mathbf{r})} \propto \mathbf{b}(\mathbf{r}),
\end{eqnarray}
in agreement with the result that follows from the standard form of the core
energy in Eq. (\ref{Hamiltonian}). Hence the degree of anisotropy in
the core energy of a hexagonal lattice is expected be a function of
the coarse graining size, with the limiting behavior being that of an isotropic system.

\subsection{Dynamics}

The equation of conservation of Burgers vector is still Eq. (\ref{cons_sq}),
but the linear response assumption relating forces and fluxes is given by
\begin{eqnarray}
\Phi_{k}^{(s)}(\mathbf{r}) = - \frac{2}{3} D^{(s)}_{kj}(\mathbf{r})
\;\; \dd_j \frac{\delta H}{\delta b^{(s)}(\mathbf{r})}. 
\label{linear}
\end{eqnarray}
The constant factor of $2/3$ corrects for the fact that the sum of the
projections onto three linearly dependent axes over represents a vector by the
factor $\gamma$ in $\sum_{s} \theta^{(s)}_k \theta^{(s)}_l = \gamma
\delta_{kl}$, which is $3/2$ for the hexagonal lattice (Eq.
(\ref{eq:hex_thetas})). 
 
The energy in terms of the Burgers vector densities is the same as in an
isotropic system. Inserting Eq. (\ref{eq:b_decomp}) into Eq.
(\ref{Hamiltonian}), we have
\begin{equation}
H_{\textrm{int}} = -\frac{K}{2} \int d\mathbf{r} d \mathbf{r}'
\sum_{s,r} \left[ \bm{\hat{\theta}}^{(s)}(\mathbf{r}) \cdot 
\bm{\hat{\theta}}^{(r)}(\mathbf{r'}) \ln\left(\frac{\rho}{a}\right)- 
\left(\bm{\hat{\theta}}^{(s)}(\mathbf{r}) \cdot \bm{\hat{\rho}}
\right) \left(\bm{\hat{\theta}}^{(r)}(\mathbf{r'}) \cdot
  \bm{\hat{\rho}} 
\right) \right]  b^{(s)}(\mathbf{r}) b^{(r)}(\mathbf{r}'),
\label{fullHam}
\end{equation}
so that the functional derivatives are
\begin{equation}
\frac{\delta H_{\textrm{int}}}{\delta b^{(s)}(\mathbf{r})} =
-K \int d \mathbf{r}' \sum_{r}
\left[\bm{\hat{\theta}}^{(s)}(\mathbf{r}) \cdot
  \bm{\hat{\theta}}^{(r)}(\mathbf{r'})
  \ln\left(\frac{\rho}{a}\right) -
 \left(\bm{\hat{\theta}}^{(s)}(\mathbf{r}) \cdot
    \bm{\hat{\rho}} \right)
  \left(\bm{\hat{\theta}}^{(r)}(\mathbf{r'}) \cdot \bm{\hat{\rho}}
  \right) \right]b^{(r)}(\mathbf{r}')
\end{equation}
or,
\begin{equation}  
\frac{\delta H_{\textrm{int}}}{\delta b^{(s)}(\mathbf{r})} = -K
\int d \mathbf{r}' \; \left[{\theta}_k^{(s)}(\mathbf{r})
  \ln\left(\frac{\rho}{a}\right) -
  \left(\bm{\hat{\theta}^{(s)}}(\mathbf{r}) \cdot \bm{\hat{\rho}}
  \right) \rho_k \right] b_k(\mathbf{r}'). 
\label{funcderiv}
\end{equation}
We now assume that the mobility in Eq. (\ref{linear}) can be
decomposed along the local slip planes into glide and climb components as in
Eq. (\ref{coefficients}). The resulting equation of motion for a
hexagonal lattice is also
\begin{equation}
\frac{\partial b^{(s)}(\mathbf{r})}{\partial t} = \frac{2}{3}
\left[(D_g - D_c) \partial_{k} \left({{\theta}}_k^{(s)}(\mathbf{r})
    {{\theta}}_l^{(s)}(\mathbf{r}) \right) \partial_l + D_c
  \nabla^2 \right] \frac{\delta H_{\textrm{int}}}{\delta
  b^{(s)}(\mathbf{r})}. 
\label{EOM}
\end{equation}
The orientations
$\bm{\hat{\theta}^{(s)}}$ follow from Eq. (\ref{theta}) and (\ref{orientation}).

If we consider again the limiting case in which the orientation is
taken to be uniform, 
$H_{\textrm{int}}$ and the kinetic equation only depend on the
Burgers vector density not on the separate components along the slip systems. This
can be shown by multiplying Eq. (\ref{EOM}) by
$\theta^{(s)}_{k}$ and summing over $s$. For rank two and four tensors the
corresponding sums are isotropic tensors due to the hexagonal lattice symmetry,
\begin{eqnarray}
\label{eq:hex_thetas}
\sum_{s} \theta^{(s)}_k \theta^{(s)}_l \theta^{(s)}_m
\theta^{(s)}_n & = & \frac{3}{8}\left[ \delta_{kl}\delta_{mn} +
  \delta_{kn}\delta_{lm} + \delta_{km}\delta_{ln} \right]   \nonumber \\
\sum_{s} \theta^{(s)}_k \theta^{(s)}_l & = & \frac{3}{2}\delta_{kl}.
\end{eqnarray}
We find,
\begin{equation}
\frac{\partial b_k(\mathbf{r})}{\partial t} 
= D_{klmn} \partial_l \partial_m \frac{\delta H_{\textrm{int}}}{\delta b_n(\mathbf{r})}, 
\label{iso}
\end{equation}
with
\begin{equation}
D_{klmn} = \frac{1}{4}\left[(D_g + 3 D_c)\delta_{lm}\delta_{kn} + (D_g
  - D_c) 2\delta_{k(l} \delta_{m)n} \right].
\label{eq:iso_mobility}
\end{equation}
This expression reduces to the expected isotropic limit of $D_g = D_c$.

Before addressing the contribution to dislocation motion in a
hexagonal lattice that arises from $H_{loc}$ , we compare our results,
Eqs. (\ref{iso}) and (\ref{eq:iso_mobility}), with prior coarse
grained treatments of the form (\ref{eq:burger_evolution}),
\cite{re:rickman97,re:limkumnerd06,re:limkumnerd07}. Comparison of
Eqs. (\ref{eq:burger_evolution}) and (\ref{iso}) leads to the
identification,
\begin{equation}
B_{jksn} = \epsilon_{jl} \epsilon_{sm} D_{klmn}
\end{equation}
where we have made repeated use of the identity
$\epsilon_{ij}\epsilon_{jn} = - \delta_{in}$ in two
dimensions. Explicit substitution of Eq. (\ref{eq:iso_mobility}) leads to, 
\begin{eqnarray}
B_{jksn} & = & \frac{1}{4} \left[ (3D_g + D_c)\delta_{js} \delta_{kn}
  + (D_c-D_g) 2\delta_{k(s} \delta_{j)n} \right] \nonumber \\ 
& = & \frac{1}{2} (D_g + D_c) \left(\delta_{j(s} \delta_{ n)k} -
  \frac{1}{2} \delta_{jk} \delta_{sn} \right) + D_g \delta_{j[s}
\delta_{ n]k}  
+ D_c \frac{1}{2} \delta_{jk} \delta_{sn} . 
\label{eq:bandd}  
\end{eqnarray}
where we have separated the tensor $B$ into
a symmetric but traceless part, an anti symmetric part, and the trace part
with respect to the first two indices (equivalently the last
two). This allows us to make a connection with the properties
of the dislocation density current $J_{jk} \propto B_{jksn}$
\cite{re:rickman97} as already argued by Limkumnerd based on volume
change arguments \cite{re:limkumnerd07}: the trace of the current is proportional to
$D_c$ so that the dislocation current is indeed traceless if there is no climb. 

%
%

Equation (\ref{eq:bandd}) is symmetric under the exchange of first and
second pairs of indices ($(j,k) \leftrightarrow (s,n)$) which is consistent
with Onsager's reciprocity relation because the rate of free energy 
change is
\begin{equation*}
\frac{d F}{d t} = - \int d^3 r \frac{\delta H}{\delta b_j} B_{mjst} 
\epsilon_{lm} \epsilon_{sb} \partial_l \partial_b \frac{\delta H}{\delta b_t}.
\end{equation*}
However, we obtain two additional allowed terms in $B_{mjst}$ compared to 
Ref. \cite{re:rickman97}. The latter only give the traceless symmetric
contribution in Eq. (\ref{eq:bandd}). In addition, and unlike prior
work, our expression for the mobility does explicitly distinguish
between climb and glide motion.

We turn next to the calculation of the contribution to the motion of
the Burgers vector density arising from the local part of the
energy. Replace $H_{\rm int}$ by $H_{loc}$ given in
Eq. (\ref{eq:core_alpha}) in Eq. (\ref{EOM}). The isotropic term in Eq. (\ref{EOM}),
proportional to $D_{c}$, leads to diffusion of
$\mathbf{b}(\mathbf{r})$ when the equation is multiplied by
$\bm{\hat{\theta}}^{(\alpha)}$ and summed over $\alpha$. However, the
term involving the longitudinal derivative is quite nontrivial because
it involves products of three bond unit vectors. When such a product
is summed over bonds $\alpha$ the resulting rank three tensor is not
independent of the bond orientation $\theta(\mathbf{r})$ although it
does have hexagonal symmetry with respect to the bond
angle. Neglecting terms involving time derivatives of the slip line
orientations, we find
the core contribution to the evolution equation to be,
\begin{equation}
\left( \frac{\partial b_{k}(\mathbf{r})}{\partial t} \right)_{c} \approx 
\frac{2}{3}(D_g-D_c) \sum_\alpha \theta^{(\alpha)}_k \theta^{(\alpha)}_l 
\theta^{(\alpha)}_m \partial_l \partial_m \frac{\delta H_{c}}{\delta 
b^{(\alpha)}(\mathbf{r})} +  \frac{2}{3} D_c \nabla^2 b_k .
\label{eq:bcore_dynamic}
\end{equation}
Given Eqs. (\ref{eq:core_alpha}) and (\ref{triplet}) we find
\begin{eqnarray}
\frac{\delta H_{c}}{\delta b^{(\alpha)}(\mathbf{r})} = E_{c} 
b^{(\alpha)}(\mathbf{r}) = E_{c} \left( t + 
\textrm{Proj}^{(\alpha)}(\mathbf{b}) \right) .
\end{eqnarray}
We define the third rank tensor
\begin{equation}
g_{klm}(\theta) =  \frac{4}{3}\sum_\alpha \theta^{(\alpha)}_k
\theta^{(\alpha)}_l \theta^{(\alpha)}_m,
\end{equation}
so that
\begin{eqnarray}
g_{xxx} & = & \cos 3\theta \quad \quad \quad g_{xxy} = \sin 3\theta \nonumber \\
g_{yyy} & = & -\sin 3\theta \quad \quad g_{yyx} = -\cos 3\theta.
\end{eqnarray}
Here again the tensor does not depend on the order of its indices since it comes
from a tensor product over a single vector. Then Eq. (\ref{eq:bcore_dynamic})
reduces to, 
\begin{eqnarray}
\left( \frac{\partial b_{k}(\mathbf{r})}{\partial t} \right)_{c} =
\frac{2}{3} D_c \nabla^2 b_k + \frac{1}{2}E_{c} (D_g-D_c)
g_{klm}(\theta) \partial_l \partial_m t(\mathbf{r}) + \ldots
\label{eq:tdyn}
\end{eqnarray}
We have not explicitly written here the term involving projections of
$\mathbf{b}$ since we just want to point out that there exists a
dependence of the motion of the Burgers vector density on geometrically
unnecessary defects through the triplet density $t$. Although
expression (\ref{eq:tdyn}) is largely formal, it does show a kinetic
equation that explicitly depends on the rotation field $\theta$
through a term that includes the density of unnecessary dislocations. 

We reiterate that the exact projections of the Burgers vector along slip
systems, and the concomitant triplet density, will depend on the size
of the coarse graining cell. As its size becomes larger, the triplet
density will decrease as the geometrically unnecessary dislocations
are averaged out. As a consequence, the contribution from
Eq. (\ref{eq:tdyn}) to the motion of the Burgers vector will
become smaller as the coarse graining cell becomes larger. Eventually,
at sufficiently long spatial scales, the evolution on the hexagonal
lattice should become the same as in an isotropic system.

In summary, although we cannot provide complete explicit equations for the
model that we have introduced except within the approximations given,
the implicit relations between magnitudes can be obtained via a
numerical implementation. In it, given the Burgers vector density
distribution as initial condition, Eq. (\ref{EOM})
needs to be iterated in time, together with Eq. (\ref{eq:tdyn}) and the related
equation that results from the local projection of $\mathbf{b}$. Equations
(\ref{theta}) and (\ref{orientation}) allow the determination of the
orientations $\bm{\hat{\theta}}^{(s)}(\mathbf{r})$ from the
densities. From the densities and the orientations the Burgers
vector follows. The interaction energy
$H_{\textrm{int}}$ can now be evaluated. Equation (\ref{triplet}) is then used
to determine the triplet density, and the system of equations evolved in time.

\section{Two edge dislocations}
\label{sec:twoedge}

The coarse grained theory presented has a simpler representation
when the defects are assumed to be discrete and isolated, although the
assumptions of the theory fail in this limit. For the sake of
illustration only, we consider in this section the motion of two point edge
dislocations and also assume that both defect interaction energies and
lattice rotation can be approximated by the results for an isotropic
solid.

Consider as initial condition two edge dislocations at $\mathbf{r} = \mathbf{r}_{1}$ and
$\mathbf{r} = \mathbf{r}_{2}$ of Burgers vectors $b \; \bm{\hat{x}}$ and
$-b \; \bm{\hat{x}}$ respectively on an undistorted, infinite, two
dimensional space. In order to avoid the complication of unnecessary
dislocations, we consider the case of only two slip planes as
would be appropriate for a square lattice. As was the case in
Sec. \ref{sec:square}, superindices correspond to
magnitudes expressed in the rotated lattice.

Insertion of these two dislocations in the otherwise undistorted
lattice leads to rotation. For the purposes of the present example, we
estimate the lattice rotation by assuming that the medium is isotropic instead,
Eq. (\ref{theta}) \cite{re:dewit73},
\begin{equation}
\theta(\mathbf{r}) = - \frac{1}{2 \pi} \left[ \frac{b (\mathbf{r} -
    \mathbf{r}_{1}) \bm{\hat{\theta}}^{0}(\mathbf{r}_{1})}{|
  \mathbf{r} - \mathbf{r}_{1} |^{2}} - \frac{b (\mathbf{r} -
    \mathbf{r}_{2}) \bm{\hat{\theta}}^{0}(\mathbf{r}_{2})}{|
  \mathbf{r} - \mathbf{r}_{2} |^{2}} \right].
\label{eq:twoedge_rotation}
\end{equation}
The initial condition (\ref{eq:twoedge_rotation}) assumes that 
the Burgers vectors are directed along one slip plane at the
location of the defects. As shown following Eq. (\ref{eq:b_decomp}),
the directions of the slip planes of this notional square lattice are 
\begin{equation} 
\bm{\hat{\theta}}^{0}(\mathbf{r}) = ( \cos \theta(\mathbf{r}), \sin
\theta(\mathbf{r}) ) \quad\quad \bm{\hat{\theta}}^{1}(\mathbf{r}) = ( \cos \theta(\mathbf{r}+\pi/2), \sin
\theta(\mathbf{r} + \pi/2) )
\label{eq:decomp_square}
\end{equation}
The rotation field (\ref{eq:twoedge_rotation}) becomes singular at the
defect location. This singularity can be eliminated, for example, by
noting that near the defect the smallest possible distance is on the
order of the lattice spacing, itself on the order of the Burgers
vector. Other models of defect structures \cite{re:lazar03} lead to
zero rotation near the defect core. We adopt the latter and by
combining Eqs. (\ref{eq:twoedge_rotation}) and
(\ref{eq:decomp_square}), we find the following implicit relations for
the lattice rotation,
\begin{eqnarray}
\theta(\mathbf{r}_{1}) & = & \frac{b}{2 \pi} \frac{(\mathbf{r}_{1} -
  \mathbf{r}_{2} )\cdot (\cos \theta(\mathbf{r}_{2}), \sin \theta (\mathbf{r}_{2}))}{|
  \mathbf{r}_{1} - \mathbf{r}_{2} |^{2}} \nonumber \\
\theta(\mathbf{r}_{2}) & = & \frac{b}{2 \pi} \frac{(\mathbf{r}_{1} -
  \mathbf{r}_{2} )\cdot (\cos \theta(\mathbf{r}_{1}), \sin \theta (\mathbf{r}_{1}))}{|
  \mathbf{r}_{1} - \mathbf{r}_{2} |^{2}}
\label{eq:rotation_initial}
\end{eqnarray}

The location of the defects and the two rotations of
Eq. (\ref{eq:rotation_initial}) constitute the initial conditions of
the problem.

\subsection{Dynamics}

Defect motion on a square lattice is governed by
Eq. (\ref{EOM_square}). The initial discrete Burgers vector distribution can
be written as 
\begin{equation}
\mathbf{b} = b \; \delta(\mathbf{r} - \mathbf{r}_{1}(t)) \;
\bm{\hat{\theta}}^{0} (\mathbf{r}_{1}) - b \; \delta(\mathbf{r} - \mathbf{r}_{2}(t)) \;
\bm{\hat{\theta}}^{0} (\mathbf{r}_{2}),
\label{eq:twoedge_burgers}
\end{equation}
where $\delta(\mathbf{r})$ is the two dimensional Dirac delta
distribution.  The initial Burgers vector of both dislocations is taken along
$\bm{\hat{\theta}}^{0}$, and given the assumed separate conservation
of Burgers vector components along each slip plane, $\mathbf{b}$ will remain
along $\bm{\hat{\theta}}^{0}$ for all times.

Given the relation $\partial_{t} \delta( \mathbf{r} - \mathbf{r}(t)) =
- \partial_{j} \left( \delta (\mathbf{r} - \mathbf{r}(t))
  \frac{dr_{j}}{dt} \right)$, the conservation law of Burgers vector,
Eq. (\ref{cons_sq}) can be written as,
\begin{equation}
b \left[  \delta (\mathbf{r} - \mathbf{r}_{1} (t)) \frac{d
    (\mathbf{r}_{1})_{k}}{dt} -  \delta (\mathbf{r} - \mathbf{r}_{2} (t)) \frac{d
    (\mathbf{r}_{2})_{k}}{dt}  \right ] = - \left( D_{g} \theta^{0}_{k}
  \theta^{0}_{j} + D_{c} ( \delta_{kj} - \theta^{0}_{k}
  \theta^{0}_{j}) \right) \partial_{j} \frac{\delta H}{\delta b^{1}(\mathbf{r})},
\label{eq:twoedge_kinetic}
\end{equation}
where we have also used the linear constitutive assumption of
Eq. (\ref{linear_square}), and the relation for the anisotropic
diffusivity of Eq. (\ref{coefficients}).

In order to compute the thermodynamic driving force in the right hand
side of Eq. (\ref{eq:twoedge_kinetic}) we write Eq. (\ref{fullHam}) as
\begin{equation}
H = - \frac{1}{2} \int d \mathbf{r} d \mathbf{r}'
V_{st}(\mathbf{r},\mathbf{r}') b^{s}(\mathbf{r}) b^{t}(\mathbf{r}').
\end{equation}
Then, given the discrete Burgers vector distribution of
Eq. (\ref{eq:twoedge_burgers}), we find
\begin{equation}
\partial_{j} \frac{\delta H}{\delta b^{1}(\mathbf{r})} = - b
\left[ \partial_{j} V(\mathbf{r},\mathbf{r}_{1}(t)) - \partial_{j}
  V(\mathbf{r},\mathbf{r}_{2}(t)) \right],
\end{equation}
also having defined $V=V_{11}$. Finally, the kinetic equation for the
location of dislocation one is
\begin{equation}
\frac{d (\mathbf{r}_{1})_{k}}{dt} =  \left( D_{g} \theta^{0}_{k}(\mathbf{r}_{1})
  \theta^{0}_{j}(\mathbf{r}_{1}) + D_{c} ( \delta_{kj} - \theta^{0}_{k}(\mathbf{r}_{1})
  \theta^{0}_{j}(\mathbf{r}_{1})) \right)
\left( -\partial_{j}V(\mathbf{r},\mathbf{r}_{2})
\right)_{\mathbf{r}=\mathbf{r}_{1}},
\label{eq:twoedge_dynamic}
\end{equation}
and the analogous equation for the second dislocation. This equation
has the form $\frac{d (\mathbf{r}_{1})_{k}}{dt} = L_{kj}F_{j}$
according to which the defect velocity equals a mobility times a
thermodynamic force. The mobility in this example depends explicitly
on lattice variables: the orientation of the slip systems at the
defect location.

The thermodynamic force can now be evaluated explicitly from the
interaction energy (\ref{fullHam}) if we also approximate it by
that of an isotropic medium. In Fourier space, it is given by
\begin{equation}
H = \frac{K}{2} \int \frac{d^{2}\mathbf{q}}{(2 \pi)^{2}}
\frac{1}{q^{2}} \left( \delta_{ij} - \hat{q}_{i}\hat{q}_{j} \right)
b_{i}(\mathbf{q}) b_{j} (-\mathbf{q}).
\end{equation}
For example, for a single edge dislocation at the origin, 
$b_{j}(\mathbf{r}) = b \delta(x_{1})\delta(x_{2}) \delta_{j1}$, the
energy of the configuration $V_{1}$ is 
\begin{equation}
V_{1} = \frac{Kb^{2}}{2} \int \frac{d^{2}\mathbf{q}}{(2 \pi)^{2}}
\frac{q_{2}^{2}}{q^{4}}.
\end{equation}
If we now consider instead two edge dislocations as in
Eq. (\ref{eq:twoedge_burgers}) with $\bm{\rho} =
\mathbf{r}_{1} - \mathbf{r}_{2}$,
their interaction energy (excluding self energies) is
\begin{equation}
V (\bm{\rho}) = - K b^{2} \int \frac{d^{2} \mathbf{q}}{(2
  \pi)^{2}} \frac{q_{2}^{2} e^{i \mathbf{q} \cdot \bm{\rho}}}{q^{4}}.
\end{equation} 
This integral can be evaluated explicitly. Let
\begin{equation}
J(\bm{\rho}) = \int d \mathbf{q} \frac{e^{i \mathbf{q} \cdot \bm{\rho}}}{q^{4}},
\label{eq:j_a}
\end{equation}
then $V (\bm{\rho}) = (Kb^{2}/(2 \pi)^{2}) (\partial^{2}  J(\bm{\rho})
/ \partial \rho_{2}^{2})$.
The two dimensional Green's function of the biharmonic operator
$\nabla^{4} G(\mathbf{r}-\mathbf{r}') = -
\delta(\mathbf{r}-\mathbf{r}')$ is $ G(\mathbf{r}-\mathbf{r}') = -
\frac{| \mathbf{r} - \mathbf{r}'|}{8 \pi} \left( \ln | \mathbf{r} -
  \mathbf{r}'| - 1 \right)$. Then $J(\bm{\rho}) = (\pi/4) | \bm{\rho}
  |^{2} \left( \ln |\bm{\rho}|^{2} -2 \right)$ and,
\begin{equation}
V(\bm{\rho}) = - \frac{Kb^{2}}{(2 \pi)^{2}} \frac{\pi}{2} \left[ 1 -
\frac{2 \rho_{2}^{2}}{\rho^{2}} - \ln \rho^{2} \right]. 
\end{equation} 
This leads to the thermodynamic forces,
\begin{equation}
- \frac{\partial V}{\partial \rho_{1}} =  - \frac{Kb^{2}}{(2 \pi)^{2}}
\frac{1}{\rho} \cos \phi \cos 2 \phi, \quad - \frac{\partial V}{\partial
  \rho_{2}} = - \frac{Kb^{2}}{(2 \pi)^{2}} \frac{1}{\rho} \sin \phi (1
+ 2 \cos^{2} \phi),
\label{eq:forces}
\end{equation}
where $\phi$ is the angle between the line joining the two
dislocations and the $x$ axis. The functional dependence in
Eq. (\ref{eq:forces}) agrees with the classical result for the
interaction force between two straight edge dislocations in an
isotropic medium (noting, e.g., that $\sin \phi (1+ 2
\cos^{2} \phi) = y(3x^{2}+y^{2})/(x^{2}+y^{2})^{3/2})$
\cite{re:landau70}. The coefficients differ in the planar
strain considered there because the stress in the direction along the
dislocation line in three dimensions is not zero, but rather
$\sigma_{33} = \nu ( \sigma_{11} + \sigma_{22})$, with $\nu$ the
Poisson ratio. This component of the stress tensor does not appear in the
purely two dimensional calculation addressed here.

Given an initial configuration comprising two edge dislocations,
Eqs. (\ref{eq:forces}) would give the force acting on each one that is
required in the right hand side of Eq. (\ref{eq:twoedge_dynamic}). The
anisotropic mobility depends of lattice rotation at the location of
each defect, which is given by
Eqs. (\ref{eq:rotation_initial}). Equation (\ref{eq:twoedge_dynamic}) then
gives the defect velocities.

\begin{figure}
\centering
\includegraphics[width=10cm]{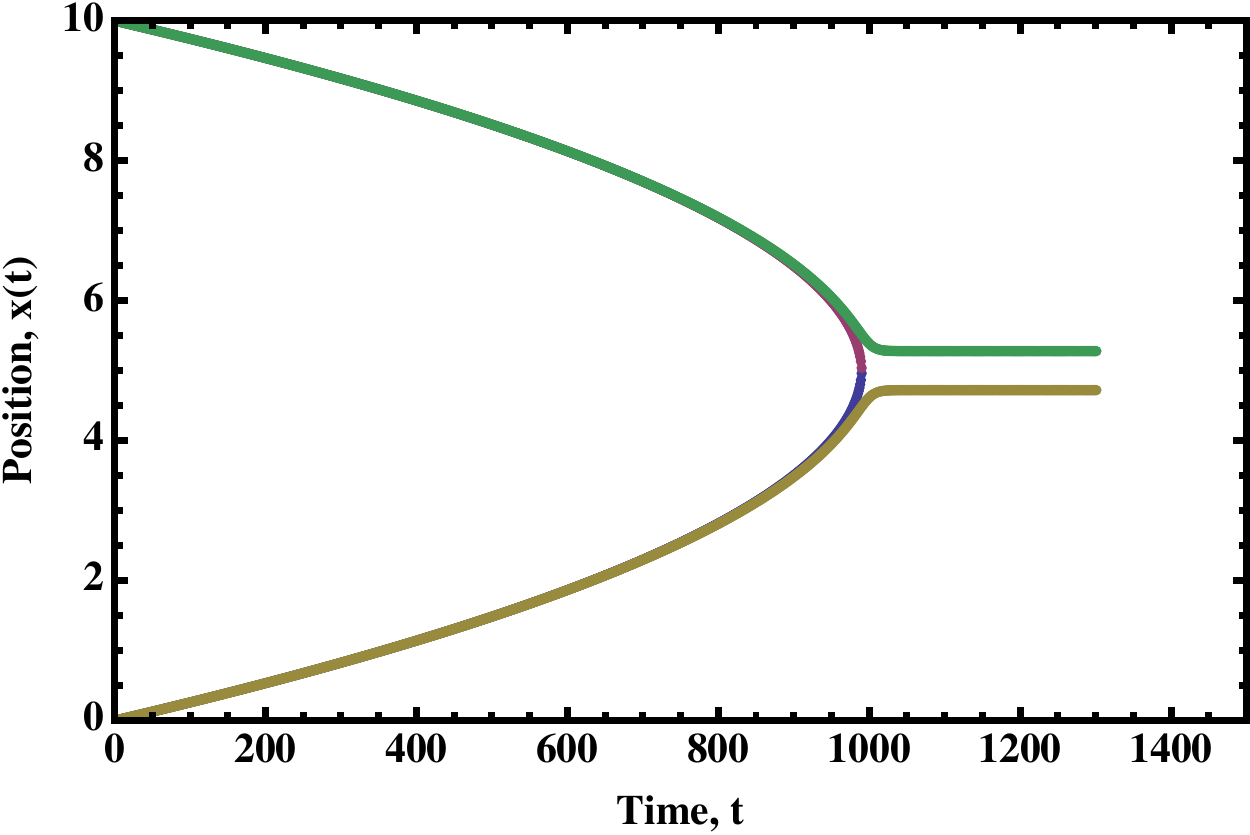}
\includegraphics[width=10cm]{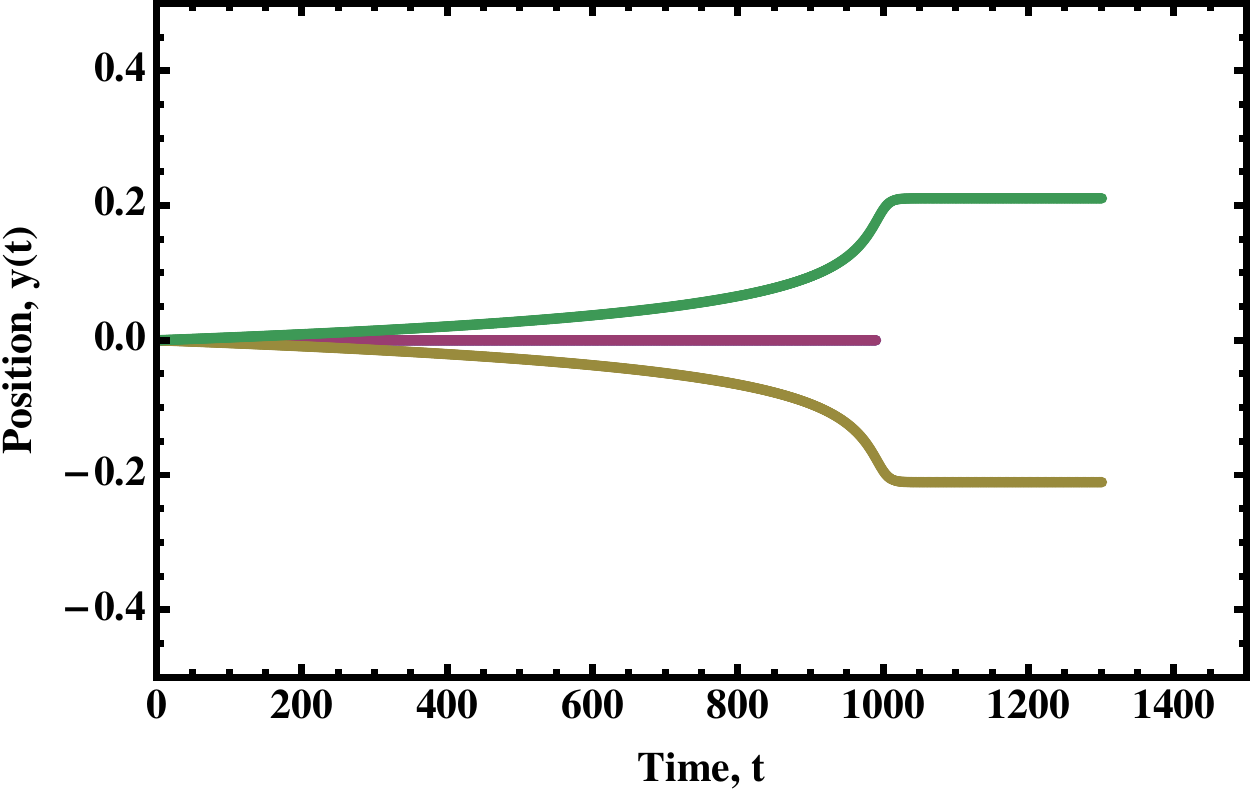}
\caption{a.) The dislocation position on the $y=0$ line, $x \left(t \right)$, as a 
function of time, $t$, for two opposite edge dislocations that are 
initially separated by a distance of $10$, in units of $b$. For each 
case there is no climb mobility. The blue 
and red curves are the positions in the absence of lattice rotations, 
while the gold and green curves pertain to a system with lattice 
rotations.  Note that, in the latter case, motion is arrested after 
some time. b.) The  
corresponding dislocation position, $y \left(t \right)$, as a 
function of time, $t$, for the two dislocations.} 
\label{fig:noclimb}
\end{figure}

We next evaluate the system of equations numerically.
Consider that the two opposite edge dislocations lie along the line
$y=0$ separated by a  
distance $10 b$.  For convenience, we work in reduced units such that 
distances are expressed in units of $b$ and speed in units of $D_{g}Kb$.  
In a first scenario, we suppose that only glide is possible 
(i.e., $D_{c}/D_{g}=0$) and examine the impact of lattice rotation 
on the motion of the dislocations.  Figure \ref{fig:noclimb}a shows the dependence 
of dislocation position on the $y=0$ line, $x \left(t \right)$, on 
time, $t$, for each dislocation.  As is evident from the 
figure, in the absence of lattice rotations, the dislocations move with increasing speed until annihilation.  Figure \ref{fig:noclimb}b shows the dislocation position normal to this plane, $y \left(t \right)$. In the absence of lattice rotations, there
is no motion perpendicular to the plane, as expected.

If lattice rotations are incorporated in the 
model, qualitatively new behavior is observed.  Figures
\ref{fig:noclimb}a and \ref{fig:noclimb}b also show that the motion of
two dislocations is similar to the case of no rotation for large
separations, but the two defects come to rest at a fixed separation. The local 
rotation of the lattice has evidently resulted in motion in the $y$-direction 
leading to the formation of a stable, dipolar configuration oriented at somewhat less 
than $45^{\circ}$ from the $x$-axis. (One would expect a $45^{\circ}$-dipole for 
two opposite edge dislocations moving on parallel slip planes in the 
absence of lattice rotations.)  

\begin{figure}
\centering
\includegraphics[width=10cm]{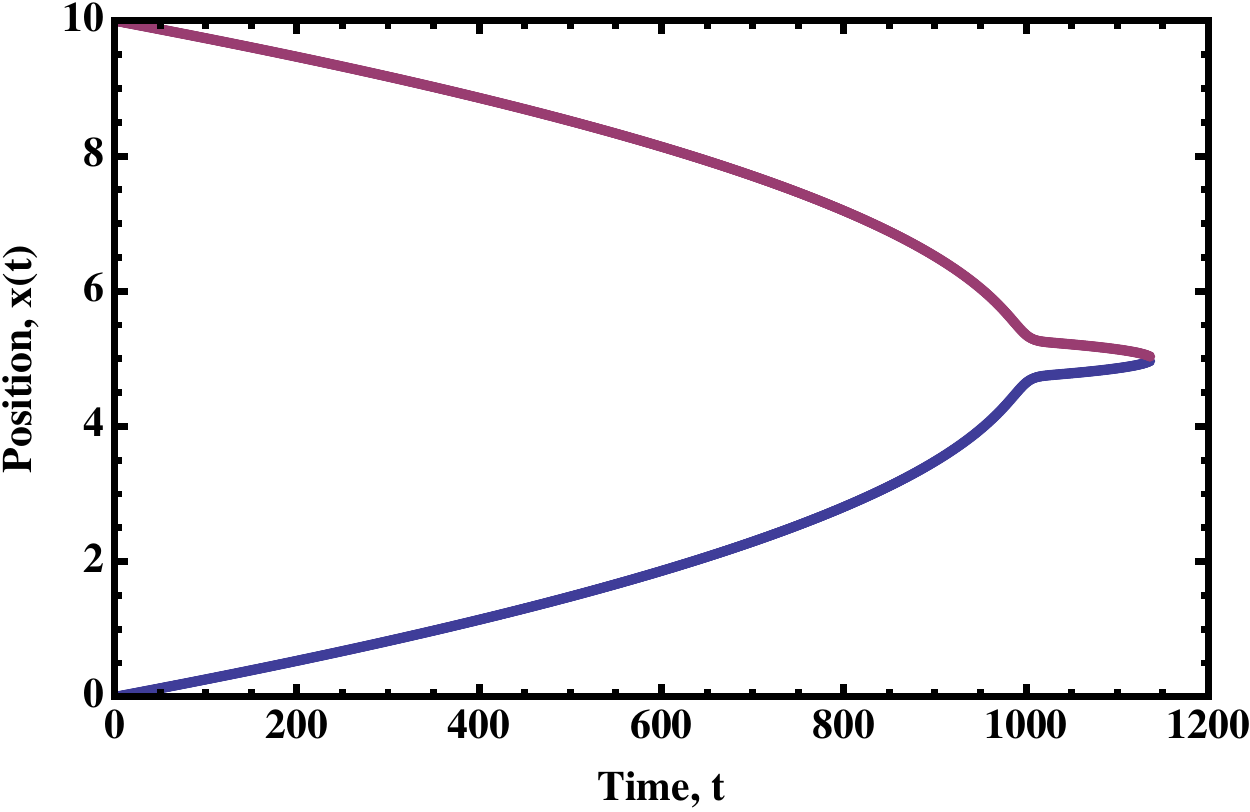}
\includegraphics[width=10cm]{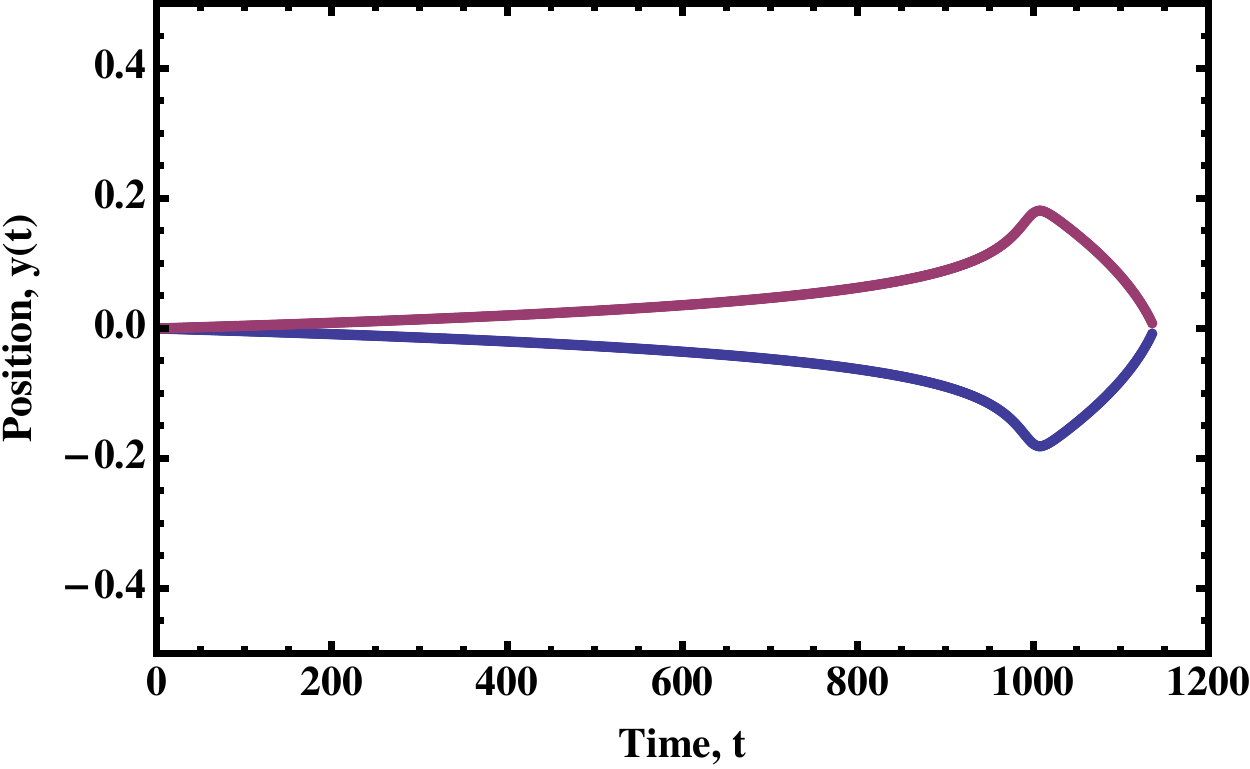}
\caption{a.) The dislocation position, $x \left(t \right)$, as a 
function of time, $t$, for the case in which both lattice rotations 
and climb are operative.  In this case, $D_{c}/D_{g}=0.02$.  The 
dipolar configuration seen in the previous figure is unstable and 
annihilation results.  b.) The corresponding dislocation position, $y \left(t \right)$, as a function of time, $t$.} 
\label{fig:climb}
\end{figure}

In the second scenario, we assess the effect of defect climb on the trajectory of the dislocations.  Figures \ref{fig:climb} 
show the positions, $x \left(t \right)$, and $y \left(t \right)$, respectively, 
for $D_{c}/D_{g}=0.02$ with lattice rotations for 
the two dislocations.  The inclusion of climb is seen to lead to an 
instability in the dipolar configuration resulting in annihilation, 
as might be expected from the functional form of the force in the 
$y$-direction given in Eqs. (\ref{eq:forces}).

This illustrative example, while highlighting the main elements and
dependencies of the model
described in Sections \ref{sec:square} and \ref{sec:hexagonal}, has
several shortcomings. First and foremost, the theory as presented is
expected to apply to a
coarse grained defect distribution but not to isolated defects. Thus the
overdamped nature of Eq. (\ref{linear_square}) can only be assumed at
the mesoscale, not at the scale of individual dislocations. Second, and
for the purpose of the example, we have used isotropic results to
compute interaction defect energies and lattice rotation, while
retaining motion along two privileged slip axes. The results of
Secs. \ref{sec:square} and \ref{sec:hexagonal} are free of these
limitations, but are considerably more involved, necessitating a fully
numerical approach for their analysis.

We conclude by mentioning that we expect that the methods described above
provide a first step into incorporating kinetic lattice effects into
continuum (coarse-grained) descriptions of defect motion. We have done
so by allowing directional defect mobilities along distinguished slip systems
in weakly distorted systems. The case of a square lattice is somewhat
simpler as the number of slip systems equals the dimensionality of the
lattice. For a hexagonal lattice, on the other hand, linear elasticity
is that of an isotropic system -a simplification- whereas the
coarse-grained model requires the introduction of geometrically
unnecessary dislocations -a complication. In this latetr case, and for large
coarse-graining volumes, a fully isotropic theory is expected albeit with separate
climb and glide diffusivities. Unfortunately, the governing equations
which we have obtained are quite complex and need to be evaluated
numerically. Such a numerical solution could be compared to direct coarse-graining of Molecular
Dynamics simulations of two-dimensional lattices. Alternatively, our
results can be verified against numerical solutions of Phase Field
Crystal models which hold at the same level of coarse graining as our theory.
Finally, for the simple example of two point edge dislocations that we
have described in
Sec. \ref{sec:twoedge}, we have shown dynamical arrest in dislocation
motion that arises from mismatches in the local
slip planes as the defects approach each other. Such an effect is
absent in a purely continuum theory.

\begin{acknowledgments}
This research has been supported by the National Science Foundation
under contract DMS 1435372 and the Minnesota Supercomputing Institute.
We thank Noah Mitchell and Zhi-Feng Huang for many useful and
stimulating discussions.
\end{acknowledgments}

\bibliographystyle{apsrev4-1}
\bibliography{refs}

\begin{thebibliography}{43}%
\makeatletter
\providecommand \@ifxundefined [1]{%
 \@ifx{#1\undefined}
}%
\providecommand \@ifnum [1]{%
 \ifnum #1\expandafter \@firstoftwo
 \else \expandafter \@secondoftwo
 \fi
}%
\providecommand \@ifx [1]{%
 \ifx #1\expandafter \@firstoftwo
 \else \expandafter \@secondoftwo
 \fi
}%
\providecommand \natexlab [1]{#1}%
\providecommand \enquote  [1]{``#1''}%
\providecommand \bibnamefont  [1]{#1}%
\providecommand \bibfnamefont [1]{#1}%
\providecommand \citenamefont [1]{#1}%
\providecommand \href@noop [0]{\@secondoftwo}%
\providecommand \href [0]{\begingroup \@sanitize@url \@href}%
\providecommand \@href[1]{\@@startlink{#1}\@@href}%
\providecommand \@@href[1]{\endgroup#1\@@endlink}%
\providecommand \@sanitize@url [0]{\catcode `\\12\catcode `\$12\catcode
  `\&12\catcode `\#12\catcode `\^12\catcode `\_12\catcode `\%12\relax}%
\providecommand \@@startlink[1]{}%
\providecommand \@@endlink[0]{}%
\providecommand \url  [0]{\begingroup\@sanitize@url \@url }%
\providecommand \@url [1]{\endgroup\@href {#1}{\urlprefix }}%
\providecommand \urlprefix  [0]{URL }%
\providecommand \Eprint [0]{\href }%
\providecommand \doibase [0]{http://dx.doi.org/}%
\providecommand \selectlanguage [0]{\@gobble}%
\providecommand \bibinfo  [0]{\@secondoftwo}%
\providecommand \bibfield  [0]{\@secondoftwo}%
\providecommand \translation [1]{[#1]}%
\providecommand \BibitemOpen [0]{}%
\providecommand \bibitemStop [0]{}%
\providecommand \bibitemNoStop [0]{.\EOS\space}%
\providecommand \EOS [0]{\spacefactor3000\relax}%
\providecommand \BibitemShut  [1]{\csname bibitem#1\endcsname}%
\let\auto@bib@innerbib\@empty
\bibitem [{\citenamefont {Nye}(1953)}]{re:nye53}%
  \BibitemOpen
  \bibfield  {author} {\bibinfo {author} {\bibfnamefont {J.}~\bibnamefont
  {Nye}},\ }\href@noop {} {\bibfield  {journal} {\bibinfo  {journal} {Acta
  metall.}\ }\textbf {\bibinfo {volume} {1}},\ \bibinfo {pages} {153} (\bibinfo
  {year} {1953})}\BibitemShut {NoStop}%
\bibitem [{\citenamefont {Kosevich}(1979)}]{re:kosevich79}%
  \BibitemOpen
  \bibfield  {author} {\bibinfo {author} {\bibfnamefont {A.}~\bibnamefont
  {Kosevich}},\ }in\ \href@noop {} {\emph {\bibinfo {booktitle} {Dislocations
  in Solids}}},\ Vol.~\bibinfo {volume} {1},\ \bibinfo {editor} {edited by\
  \bibinfo {editor} {\bibfnamefont {F.}~\bibnamefont {Nabarro}}}\ (\bibinfo
  {publisher} {North-Holland},\ \bibinfo {address} {New York},\ \bibinfo {year}
  {1979})\ p.~\bibinfo {pages} {33}\BibitemShut {NoStop}%
\bibitem [{\citenamefont {{Kr\"oner}}(1981)}]{re:kroner81}%
  \BibitemOpen
  \bibfield  {author} {\bibinfo {author} {\bibfnamefont {E.}~\bibnamefont
  {{Kr\"oner}}},\ }in\ \href@noop {} {\emph {\bibinfo {booktitle} {Physics of
  Defects}}},\ \bibinfo {editor} {edited by\ \bibinfo {editor} {\bibfnamefont
  {R.}~\bibnamefont {Balian}}, \bibinfo {editor} {\bibfnamefont
  {M.}~\bibnamefont {{Kl\'eman}}}, \ and\ \bibinfo {editor} {\bibfnamefont
  {J.}~\bibnamefont {Poirier}}}\ (\bibinfo  {publisher} {North Holland},\
  \bibinfo {year} {1981})\ \bibinfo {note} {proceedings of Les Houches, Session
  XXXV}\BibitemShut {NoStop}%
\bibitem [{\citenamefont {Mura}(1987)}]{re:mura87}%
  \BibitemOpen
  \bibfield  {author} {\bibinfo {author} {\bibfnamefont {T.}~\bibnamefont
  {Mura}},\ }\href@noop {} {\emph {\bibinfo {title} {Micromechanics of Defects
  in Solids}}}\ (\bibinfo  {publisher} {Martinus Nijhoff},\ \bibinfo {address}
  {Boston},\ \bibinfo {year} {1987})\BibitemShut {NoStop}%
\bibitem [{\citenamefont {Nelson}(2002)}]{re:nelson02}%
  \BibitemOpen
  \bibfield  {author} {\bibinfo {author} {\bibfnamefont {D.}~\bibnamefont
  {Nelson}},\ }\href@noop {} {\emph {\bibinfo {title} {Defects and Geometry in
  Condensed Matter Physics}}}\ (\bibinfo  {publisher} {Cambridge University
  Press},\ \bibinfo {address} {New York},\ \bibinfo {year} {2002})\BibitemShut
  {NoStop}%
\bibitem [{\citenamefont {Arsenlis}\ and\ \citenamefont
  {Parks}(1999)}]{re:arsemlis99}%
  \BibitemOpen
  \bibfield  {author} {\bibinfo {author} {\bibfnamefont {A.}~\bibnamefont
  {Arsenlis}}\ and\ \bibinfo {author} {\bibfnamefont {D.}~\bibnamefont
  {Parks}},\ }\href@noop {} {\bibfield  {journal} {\bibinfo  {journal} {Acta
  mater.}\ }\textbf {\bibinfo {volume} {47}},\ \bibinfo {pages} {1597}
  (\bibinfo {year} {1999})}\BibitemShut {NoStop}%
\bibitem [{\citenamefont {Nelson}\ and\ \citenamefont
  {Halperin}(1979)}]{re:nelson79}%
  \BibitemOpen
  \bibfield  {author} {\bibinfo {author} {\bibfnamefont {D.~R.}\ \bibnamefont
  {Nelson}}\ and\ \bibinfo {author} {\bibfnamefont {B.~I.}\ \bibnamefont
  {Halperin}},\ }\href {\doibase 10.1103/PhysRevB.19.2457} {\bibfield
  {journal} {\bibinfo  {journal} {Phys. Rev. B}\ }\textbf {\bibinfo {volume}
  {19}},\ \bibinfo {pages} {2457} (\bibinfo {year} {1979})}\BibitemShut
  {NoStop}%
\bibitem [{\citenamefont {Zippelius}\ \emph {et~al.}(1980)\citenamefont
  {Zippelius}, \citenamefont {Halperin},\ and\ \citenamefont
  {Nelson}}]{re:zippelius80}%
  \BibitemOpen
  \bibfield  {author} {\bibinfo {author} {\bibfnamefont {A.}~\bibnamefont
  {Zippelius}}, \bibinfo {author} {\bibfnamefont {B.~I.}\ \bibnamefont
  {Halperin}}, \ and\ \bibinfo {author} {\bibfnamefont {D.~R.}\ \bibnamefont
  {Nelson}},\ }\href {\doibase 10.1103/PhysRevB.22.2514} {\bibfield  {journal}
  {\bibinfo  {journal} {Phys. Rev. B}\ }\textbf {\bibinfo {volume} {22}},\
  \bibinfo {pages} {2514} (\bibinfo {year} {1980})}\BibitemShut {NoStop}%
\bibitem [{\citenamefont {Nelson}\ and\ \citenamefont
  {Toner}(1981)}]{re:nelson81}%
  \BibitemOpen
  \bibfield  {author} {\bibinfo {author} {\bibfnamefont {D.~R.}\ \bibnamefont
  {Nelson}}\ and\ \bibinfo {author} {\bibfnamefont {J.}~\bibnamefont {Toner}},\
  }\href {\doibase 10.1103/PhysRevB.24.363} {\bibfield  {journal} {\bibinfo
  {journal} {Phys. Rev. B}\ }\textbf {\bibinfo {volume} {24}},\ \bibinfo
  {pages} {363} (\bibinfo {year} {1981})}\BibitemShut {NoStop}%
\bibitem [{\citenamefont {Groma}(1997)}]{re:groma97}%
  \BibitemOpen
  \bibfield  {author} {\bibinfo {author} {\bibfnamefont {I.}~\bibnamefont
  {Groma}},\ }\href@noop {} {\bibfield  {journal} {\bibinfo  {journal} {Phys.
  Rev. B}\ }\textbf {\bibinfo {volume} {56}},\ \bibinfo {pages} {5807}
  (\bibinfo {year} {1997})}\BibitemShut {NoStop}%
\bibitem [{\citenamefont {Rickman}\ and\ \citenamefont
  {{Vi\~nals}}(1997)}]{re:rickman97}%
  \BibitemOpen
  \bibfield  {author} {\bibinfo {author} {\bibfnamefont {J.}~\bibnamefont
  {Rickman}}\ and\ \bibinfo {author} {\bibfnamefont {J.}~\bibnamefont
  {{Vi\~nals}}},\ }\href@noop {} {\bibfield  {journal} {\bibinfo  {journal}
  {Phil. Mag. A}\ }\textbf {\bibinfo {volume} {75}},\ \bibinfo {pages} {1251}
  (\bibinfo {year} {1997})}\BibitemShut {NoStop}%
\bibitem [{\citenamefont {Zaiser}\ \emph {et~al.}(2001)\citenamefont {Zaiser},
  \citenamefont {Miguel},\ and\ \citenamefont {Groma}}]{re:zaiser01}%
  \BibitemOpen
  \bibfield  {author} {\bibinfo {author} {\bibfnamefont {M.}~\bibnamefont
  {Zaiser}}, \bibinfo {author} {\bibfnamefont {M.-C.}\ \bibnamefont {Miguel}},
  \ and\ \bibinfo {author} {\bibfnamefont {I.}~\bibnamefont {Groma}},\
  }\href@noop {} {\bibfield  {journal} {\bibinfo  {journal} {Phys. Rev. B}\
  }\textbf {\bibinfo {volume} {64}},\ \bibinfo {pages} {224102} (\bibinfo
  {year} {2001})}\BibitemShut {NoStop}%
\bibitem [{\citenamefont {Limkumnerd}\ and\ \citenamefont
  {Sethna}(2006)}]{re:limkumnerd06}%
  \BibitemOpen
  \bibfield  {author} {\bibinfo {author} {\bibfnamefont {S.}~\bibnamefont
  {Limkumnerd}}\ and\ \bibinfo {author} {\bibfnamefont {J.~P.}\ \bibnamefont
  {Sethna}},\ }\href {\doibase 10.1103/PhysRevLett.96.095503} {\bibfield
  {journal} {\bibinfo  {journal} {Phys. Rev. Lett.}\ }\textbf {\bibinfo
  {volume} {96}},\ \bibinfo {pages} {095503} (\bibinfo {year}
  {2006})}\BibitemShut {NoStop}%
\bibitem [{\citenamefont {Acharya}(2010)}]{re:acharya10}%
  \BibitemOpen
  \bibfield  {author} {\bibinfo {author} {\bibfnamefont {A.}~\bibnamefont
  {Acharya}},\ }\href@noop {} {\bibfield  {journal} {\bibinfo  {journal} {J.
  Mech. Phys. Solids}\ }\textbf {\bibinfo {volume} {58}},\ \bibinfo {pages}
  {766} (\bibinfo {year} {2010})}\BibitemShut {NoStop}%
\bibitem [{\citenamefont {Hytch}\ \emph {et~al.}(2003)\citenamefont {Hytch},
  \citenamefont {Putaux},\ and\ \citenamefont {Penisson}}]{re:hytch03}%
  \BibitemOpen
  \bibfield  {author} {\bibinfo {author} {\bibfnamefont {M.}~\bibnamefont
  {Hytch}}, \bibinfo {author} {\bibfnamefont {J.-L.}\ \bibnamefont {Putaux}}, \
  and\ \bibinfo {author} {\bibfnamefont {J.-M.}\ \bibnamefont {Penisson}},\
  }\href@noop {} {\bibfield  {journal} {\bibinfo  {journal} {Nature}\ }\textbf
  {\bibinfo {volume} {423}},\ \bibinfo {pages} {270} (\bibinfo {year}
  {2003})}\BibitemShut {NoStop}%
\bibitem [{\citenamefont {Pertsinidis}\ and\ \citenamefont
  {Ling}(2001{\natexlab{a}})}]{re:pertsinidis01}%
  \BibitemOpen
  \bibfield  {author} {\bibinfo {author} {\bibfnamefont {A.}~\bibnamefont
  {Pertsinidis}}\ and\ \bibinfo {author} {\bibfnamefont {X.}~\bibnamefont
  {Ling}},\ }\href {\doibase 10.1103/PhysRevLett.87.098303} {\bibfield
  {journal} {\bibinfo  {journal} {Phys. Rev. Lett.}\ }\textbf {\bibinfo
  {volume} {87}},\ \bibinfo {pages} {098303} (\bibinfo {year}
  {2001}{\natexlab{a}})}\BibitemShut {NoStop}%
\bibitem [{\citenamefont {Pertsinidis}\ and\ \citenamefont
  {Ling}(2001{\natexlab{b}})}]{re:pertsinidis01b}%
  \BibitemOpen
  \bibfield  {author} {\bibinfo {author} {\bibfnamefont {A.}~\bibnamefont
  {Pertsinidis}}\ and\ \bibinfo {author} {\bibfnamefont {X.}~\bibnamefont
  {Ling}},\ }\href {\doibase 10.1038/35093077} {\bibfield  {journal} {\bibinfo
  {journal} {Nature}\ }\textbf {\bibinfo {volume} {413}},\ \bibinfo {pages}
  {147} (\bibinfo {year} {2001}{\natexlab{b}})}\BibitemShut {NoStop}%
\bibitem [{\citenamefont {Irvine}\ \emph {et~al.}(2013)\citenamefont {Irvine},
  \citenamefont {Hollingsworth}, \citenamefont {Grier},\ and\ \citenamefont
  {Chaikin}}]{re:irvine13}%
  \BibitemOpen
  \bibfield  {author} {\bibinfo {author} {\bibfnamefont {W.}~\bibnamefont
  {Irvine}}, \bibinfo {author} {\bibfnamefont {A.}~\bibnamefont
  {Hollingsworth}}, \bibinfo {author} {\bibfnamefont {D.}~\bibnamefont
  {Grier}}, \ and\ \bibinfo {author} {\bibfnamefont {P.}~\bibnamefont
  {Chaikin}},\ }\href@noop {} {\bibfield  {journal} {\bibinfo  {journal} {Proc.
  Natl. Acad. Sci. USA}\ }\textbf {\bibinfo {volume} {110}},\ \bibinfo {pages}
  {15544} (\bibinfo {year} {2013})}\BibitemShut {NoStop}%
\bibitem [{\citenamefont {Zhang}\ \emph {et~al.}(2015)\citenamefont {Zhang},
  \citenamefont {Grubb}, \citenamefont {Seed}, \citenamefont {Sampson},
  \citenamefont {Jakli},\ and\ \citenamefont {Lavrentovich}}]{re:zhang15}%
  \BibitemOpen
  \bibfield  {author} {\bibinfo {author} {\bibfnamefont {C.}~\bibnamefont
  {Zhang}}, \bibinfo {author} {\bibfnamefont {A.}~\bibnamefont {Grubb}},
  \bibinfo {author} {\bibfnamefont {A.}~\bibnamefont {Seed}}, \bibinfo {author}
  {\bibfnamefont {P.}~\bibnamefont {Sampson}}, \bibinfo {author} {\bibfnamefont
  {A.}~\bibnamefont {Jakli}}, \ and\ \bibinfo {author} {\bibfnamefont
  {O.}~\bibnamefont {Lavrentovich}},\ }\href {\doibase
  10.1103/PhysRevLett.115.087801} {\bibfield  {journal} {\bibinfo  {journal}
  {Phys. Rev. Lett.}\ }\textbf {\bibinfo {volume} {115}},\ \bibinfo {pages}
  {087801} (\bibinfo {year} {2015})}\BibitemShut {NoStop}%
\bibitem [{\citenamefont {Guinea}\ \emph {et~al.}(2010)\citenamefont {Guinea},
  \citenamefont {Katsnelson},\ and\ \citenamefont {Geim}}]{re:guinea10}%
  \BibitemOpen
  \bibfield  {author} {\bibinfo {author} {\bibfnamefont {F.}~\bibnamefont
  {Guinea}}, \bibinfo {author} {\bibfnamefont {M.}~\bibnamefont {Katsnelson}},
  \ and\ \bibinfo {author} {\bibfnamefont {A.}~\bibnamefont {Geim}},\
  }\href@noop {} {\bibfield  {journal} {\bibinfo  {journal} {Nature Physics}\
  }\textbf {\bibinfo {volume} {6}},\ \bibinfo {pages} {30} (\bibinfo {year}
  {2010})}\BibitemShut {NoStop}%
\bibitem [{\citenamefont {Chen}\ and\ \citenamefont
  {Chrzan}(2011)}]{re:chen11}%
  \BibitemOpen
  \bibfield  {author} {\bibinfo {author} {\bibfnamefont {S.}~\bibnamefont
  {Chen}}\ and\ \bibinfo {author} {\bibfnamefont {D.~C.}\ \bibnamefont
  {Chrzan}},\ }\href {\doibase 10.1103/PhysRevB.84.214103} {\bibfield
  {journal} {\bibinfo  {journal} {Phys. Rev. B}\ }\textbf {\bibinfo {volume}
  {84}},\ \bibinfo {pages} {214103} (\bibinfo {year} {2011})}\BibitemShut
  {NoStop}%
\bibitem [{\citenamefont {Bonilla}\ and\ \citenamefont
  {Carpio}(2012)}]{re:bonilla12}%
  \BibitemOpen
  \bibfield  {author} {\bibinfo {author} {\bibfnamefont {L.}~\bibnamefont
  {Bonilla}}\ and\ \bibinfo {author} {\bibfnamefont {A.}~\bibnamefont
  {Carpio}},\ }\href@noop {} {\  (\bibinfo {year} {2012})},\ \bibinfo {note}
  {arXiv:cond-mat.mes-hall/1207.5675}\BibitemShut {NoStop}%
\bibitem [{\citenamefont {LeSar}\ and\ \citenamefont
  {Rickman}(2002)}]{re:lesar02}%
  \BibitemOpen
  \bibfield  {author} {\bibinfo {author} {\bibfnamefont {R.}~\bibnamefont
  {LeSar}}\ and\ \bibinfo {author} {\bibfnamefont {J.}~\bibnamefont
  {Rickman}},\ }\href@noop {} {\bibfield  {journal} {\bibinfo  {journal} {Phys.
  Rev. B}\ }\textbf {\bibinfo {volume} {65}},\ \bibinfo {pages} {144110}
  (\bibinfo {year} {2002})}\BibitemShut {NoStop}%
\bibitem [{\citenamefont {Gulluoglu}\ \emph {et~al.}(1990)\citenamefont
  {Gulluoglu}, \citenamefont {Srolovitz}, \citenamefont {LeSar},\ and\
  \citenamefont {Lomdahl}}]{re:gulluoglu90}%
  \BibitemOpen
  \bibfield  {author} {\bibinfo {author} {\bibfnamefont {A.}~\bibnamefont
  {Gulluoglu}}, \bibinfo {author} {\bibfnamefont {D.}~\bibnamefont
  {Srolovitz}}, \bibinfo {author} {\bibfnamefont {R.}~\bibnamefont {LeSar}}, \
  and\ \bibinfo {author} {\bibfnamefont {P.}~\bibnamefont {Lomdahl}},\
  }\href@noop {} {\emph {\bibinfo {title} {Simulation and Theory of Evolving
  Microstructures}}}\ (\bibinfo  {publisher} {TMS},\ \bibinfo {address}
  {Warrendale, PA},\ \bibinfo {year} {1990})\BibitemShut {NoStop}%
\bibitem [{\citenamefont {Yamakov}\ \emph {et~al.}(2002)\citenamefont
  {Yamakov}, \citenamefont {Wolf}, \citenamefont {Phillpot}, \citenamefont
  {Mukherjee},\ and\ \citenamefont {Geilter}}]{re:yamakov02}%
  \BibitemOpen
  \bibfield  {author} {\bibinfo {author} {\bibfnamefont {V.}~\bibnamefont
  {Yamakov}}, \bibinfo {author} {\bibfnamefont {D.}~\bibnamefont {Wolf}},
  \bibinfo {author} {\bibfnamefont {S.}~\bibnamefont {Phillpot}}, \bibinfo
  {author} {\bibfnamefont {A.}~\bibnamefont {Mukherjee}}, \ and\ \bibinfo
  {author} {\bibfnamefont {H.}~\bibnamefont {Geilter}},\ }\href {\doibase
  10.1038/nmat700} {\bibfield  {journal} {\bibinfo  {journal} {Nature
  Materials}\ }\textbf {\bibinfo {volume} {1}},\ \bibinfo {pages} {45}
  (\bibinfo {year} {2002})}\BibitemShut {NoStop}%
\bibitem [{\citenamefont {Crone}\ \emph {et~al.}(2014)\citenamefont {Crone},
  \citenamefont {Chung}, \citenamefont {Leiter}, \citenamefont {Knap},
  \citenamefont {Aubry}, \citenamefont {Hommes},\ and\ \citenamefont
  {Arsenlis}}]{re:crone14}%
  \BibitemOpen
  \bibfield  {author} {\bibinfo {author} {\bibfnamefont {J.}~\bibnamefont
  {Crone}}, \bibinfo {author} {\bibfnamefont {P.}~\bibnamefont {Chung}},
  \bibinfo {author} {\bibfnamefont {K.}~\bibnamefont {Leiter}}, \bibinfo
  {author} {\bibfnamefont {J.}~\bibnamefont {Knap}}, \bibinfo {author}
  {\bibfnamefont {S.}~\bibnamefont {Aubry}}, \bibinfo {author} {\bibfnamefont
  {G.}~\bibnamefont {Hommes}}, \ and\ \bibinfo {author} {\bibfnamefont
  {A.}~\bibnamefont {Arsenlis}},\ }\href {\doibase
  10.1088/0965-0393/22/3/035014} {\bibfield  {journal} {\bibinfo  {journal}
  {Modelling Simul. Mater. Sci. Eng.}\ }\textbf {\bibinfo {volume} {22}},\
  \bibinfo {pages} {035014} (\bibinfo {year} {2014})}\BibitemShut {NoStop}%
\bibitem [{\citenamefont {{Gr\"oger}}\ \emph {et~al.}(2008)\citenamefont
  {{Gr\"oger}}, \citenamefont {Lookman},\ and\ \citenamefont
  {Saxena}}]{re:groger08}%
  \BibitemOpen
  \bibfield  {author} {\bibinfo {author} {\bibfnamefont {R.}~\bibnamefont
  {{Gr\"oger}}}, \bibinfo {author} {\bibfnamefont {T.}~\bibnamefont {Lookman}},
  \ and\ \bibinfo {author} {\bibfnamefont {A.}~\bibnamefont {Saxena}},\
  }\href@noop {} {\bibfield  {journal} {\bibinfo  {journal} {Phys. Rev. B}\
  }\textbf {\bibinfo {volume} {78}},\ \bibinfo {pages} {184101} (\bibinfo
  {year} {2008})}\BibitemShut {NoStop}%
\bibitem [{\citenamefont {Li}\ \emph {et~al.}(2008)\citenamefont {Li},
  \citenamefont {Hu}, \citenamefont {Choudhury}, \citenamefont {Baskes},
  \citenamefont {Saxena}, \citenamefont {Lookman}, \citenamefont {Jia},
  \citenamefont {Schlom},\ and\ \citenamefont {Chen}}]{re:li08}%
  \BibitemOpen
  \bibfield  {author} {\bibinfo {author} {\bibfnamefont {Y.~L.}\ \bibnamefont
  {Li}}, \bibinfo {author} {\bibfnamefont {S.~Y.}\ \bibnamefont {Hu}}, \bibinfo
  {author} {\bibfnamefont {S.}~\bibnamefont {Choudhury}}, \bibinfo {author}
  {\bibfnamefont {M.~I.}\ \bibnamefont {Baskes}}, \bibinfo {author}
  {\bibfnamefont {A.}~\bibnamefont {Saxena}}, \bibinfo {author} {\bibfnamefont
  {T.}~\bibnamefont {Lookman}}, \bibinfo {author} {\bibfnamefont {Q.~X.}\
  \bibnamefont {Jia}}, \bibinfo {author} {\bibfnamefont {D.~G.}\ \bibnamefont
  {Schlom}}, \ and\ \bibinfo {author} {\bibfnamefont {L.~Q.}\ \bibnamefont
  {Chen}},\ }\href {\doibase 10.1063/1.3021354} {\bibfield  {journal} {\bibinfo
   {journal} {J. Appl. Phys.}\ }\textbf {\bibinfo {volume} {104}},\ \bibinfo
  {eid} {104110} (\bibinfo {year} {2008})}\BibitemShut {NoStop}%
\bibitem [{\citenamefont {Chen}\ \emph {et~al.}(2013)\citenamefont {Chen},
  \citenamefont {Choi}, \citenamefont {Papanikolaou}, \citenamefont
  {Bierbaum},\ and\ \citenamefont {Sethna}}]{re:chen13}%
  \BibitemOpen
  \bibfield  {author} {\bibinfo {author} {\bibfnamefont {Y.~S.}\ \bibnamefont
  {Chen}}, \bibinfo {author} {\bibfnamefont {W.}~\bibnamefont {Choi}}, \bibinfo
  {author} {\bibfnamefont {S.}~\bibnamefont {Papanikolaou}}, \bibinfo {author}
  {\bibfnamefont {M.}~\bibnamefont {Bierbaum}}, \ and\ \bibinfo {author}
  {\bibfnamefont {J.~P.}\ \bibnamefont {Sethna}},\ }\href {\doibase
  10.1016/j.ijplas.2013.02.011} {\bibfield  {journal} {\bibinfo  {journal}
  {International Journal of Plasticity}\ }\textbf {\bibinfo {volume} {46}},\
  \bibinfo {pages} {94 } (\bibinfo {year} {2013})}\BibitemShut {NoStop}%
\bibitem [{\citenamefont {Acharya}(2001)}]{re:acharya01}%
  \BibitemOpen
  \bibfield  {author} {\bibinfo {author} {\bibfnamefont {A.}~\bibnamefont
  {Acharya}},\ }\href@noop {} {\bibfield  {journal} {\bibinfo  {journal} {J.
  Mech. Phys. Solids}\ }\textbf {\bibinfo {volume} {49}},\ \bibinfo {pages}
  {761} (\bibinfo {year} {2001})}\BibitemShut {NoStop}%
\bibitem [{\citenamefont {Bowick}\ and\ \citenamefont
  {Travesset}(2001)}]{re:Bowick01}%
  \BibitemOpen
  \bibfield  {author} {\bibinfo {author} {\bibfnamefont {M.}~\bibnamefont
  {Bowick}}\ and\ \bibinfo {author} {\bibfnamefont {A.}~\bibnamefont
  {Travesset}},\ }\href {\doibase 10.1088/0305-4470/34/8/301} {\bibfield
  {journal} {\bibinfo  {journal} {J. Phys. A: Math. Gen.}\ }\textbf {\bibinfo
  {volume} {34}},\ \bibinfo {pages} {1535} (\bibinfo {year}
  {2001})}\BibitemShut {NoStop}%
\bibitem [{\citenamefont {Nelson}(1978)}]{re:Nelson78}%
  \BibitemOpen
  \bibfield  {author} {\bibinfo {author} {\bibfnamefont {D.~R.}\ \bibnamefont
  {Nelson}},\ }\href {\doibase 10.1103/PhysRevB.18.2318} {\bibfield  {journal}
  {\bibinfo  {journal} {Phys. Rev. B}\ }\textbf {\bibinfo {volume} {18}},\
  \bibinfo {pages} {2318} (\bibinfo {year} {1978})}\BibitemShut {NoStop}%
\bibitem [{\citenamefont {Chaikin}\ and\ \citenamefont
  {Lubensky}(1995)}]{re:Chaikin95}%
  \BibitemOpen
  \bibfield  {author} {\bibinfo {author} {\bibfnamefont {P.}~\bibnamefont
  {Chaikin}}\ and\ \bibinfo {author} {\bibfnamefont {T.}~\bibnamefont
  {Lubensky}},\ }\href@noop {} {\emph {\bibinfo {title} {Principles of
  condensed matter physics}}}\ (\bibinfo  {publisher} {Cambridge University
  Press},\ \bibinfo {address} {New York},\ \bibinfo {year} {1995})\BibitemShut
  {NoStop}%
\bibitem [{\citenamefont {Fisher}\ \emph {et~al.}(1979)\citenamefont {Fisher},
  \citenamefont {Halperin},\ and\ \citenamefont {Morf}}]{re:Fisher79}%
  \BibitemOpen
  \bibfield  {author} {\bibinfo {author} {\bibfnamefont {D.~S.}\ \bibnamefont
  {Fisher}}, \bibinfo {author} {\bibfnamefont {B.~I.}\ \bibnamefont
  {Halperin}}, \ and\ \bibinfo {author} {\bibfnamefont {R.}~\bibnamefont
  {Morf}},\ }\href {\doibase 10.1103/PhysRevB.20.4692} {\bibfield  {journal}
  {\bibinfo  {journal} {Phys. Rev. B}\ }\textbf {\bibinfo {volume} {20}},\
  \bibinfo {pages} {4692} (\bibinfo {year} {1979})}\BibitemShut {NoStop}%
\bibitem [{\citenamefont {Aguenaou}(1997)}]{re:aguenaou97}%
  \BibitemOpen
  \bibfield  {author} {\bibinfo {author} {\bibfnamefont {K.}~\bibnamefont
  {Aguenaou}},\ }\emph {\bibinfo {title} {Modeling of Solidification}},\
  \href@noop {} {Ph.D. thesis},\ \bibinfo  {school} {McGill University}
  (\bibinfo {year} {1997})\BibitemShut {NoStop}%
\bibitem [{\citenamefont {Haataja}\ \emph {et~al.}(2002)\citenamefont
  {Haataja}, \citenamefont {M{\"u}ller}, \citenamefont {Rutenberg},\ and\
  \citenamefont {Grant}}]{re:haataja02}%
  \BibitemOpen
  \bibfield  {author} {\bibinfo {author} {\bibfnamefont {M.}~\bibnamefont
  {Haataja}}, \bibinfo {author} {\bibfnamefont {J.}~\bibnamefont {M{\"u}ller}},
  \bibinfo {author} {\bibfnamefont {A.~D.}\ \bibnamefont {Rutenberg}}, \ and\
  \bibinfo {author} {\bibfnamefont {M.}~\bibnamefont {Grant}},\ }\href
  {\doibase 10.1103/PhysRevB.65.165414} {\bibfield  {journal} {\bibinfo
  {journal} {Phys. Rev. B}\ }\textbf {\bibinfo {volume} {65}},\ \bibinfo
  {pages} {165414} (\bibinfo {year} {2002})}\BibitemShut {NoStop}%
\bibitem [{\citenamefont {Limkumnerd}(2007)}]{re:limkumnerd07}%
  \BibitemOpen
  \bibfield  {author} {\bibinfo {author} {\bibfnamefont {S.}~\bibnamefont
  {Limkumnerd}},\ }\emph {\bibinfo {title} {Mesoscale Theory of Grains and
  Cells: Polycrystals \& Plasticity}},\ \href@noop {} {Ph.D. thesis},\ \bibinfo
   {school} {Cornell University} (\bibinfo {year} {2007})\BibitemShut {NoStop}%
\bibitem [{\citenamefont {deWit}(1973)}]{re:dewit73}%
  \BibitemOpen
  \bibfield  {author} {\bibinfo {author} {\bibfnamefont {R.}~\bibnamefont
  {deWit}},\ }\href {\doibase 10.6028/jres.077A.036} {\bibfield  {journal}
  {\bibinfo  {journal} {J. Res. NBS-A. Physics and Chemistry}\ }\textbf
  {\bibinfo {volume} {77A}},\ \bibinfo {pages} {607} (\bibinfo {year}
  {1973})}\BibitemShut {NoStop}%
\bibitem [{fo:()}]{fo:bp1_01}%
  \BibitemOpen
  \href@noop {} {}\bibinfo {note} {For a finite system the coordinate system is
  set by subtracting everywhere the calculated orientation at a reference point
  (so that the orientation vanishes at the reference point). Then the evolution
  of Burgers vector densities lead to a time dependence of the subtracted
  constant, and thus it must be kept as part of the dynamics on the right side
  of Eq. (\ref{theta}).}\BibitemShut {Stop}%
\bibitem [{\citenamefont {Steeds}(1973)}]{re:steeds73}%
  \BibitemOpen
  \bibfield  {author} {\bibinfo {author} {\bibfnamefont {J.~W.}\ \bibnamefont
  {Steeds}},\ }\href@noop {} {\emph {\bibinfo {title} {Introduction to
  anisotropic elasticity theory of dislocations}}}\ (\bibinfo  {publisher}
  {Clarendon Press},\ \bibinfo {address} {Oxford},\ \bibinfo {year}
  {1973})\BibitemShut {NoStop}%
\bibitem [{\citenamefont {Tsimring}(1995)}]{re:tsimring95}%
  \BibitemOpen
  \bibfield  {author} {\bibinfo {author} {\bibfnamefont {L.}~\bibnamefont
  {Tsimring}},\ }\href@noop {} {\bibfield  {journal} {\bibinfo  {journal}
  {Phys. Rev. Lett.}\ }\textbf {\bibinfo {volume} {74}},\ \bibinfo {pages}
  {4201} (\bibinfo {year} {1995})}\BibitemShut {NoStop}%
\bibitem [{\citenamefont {Lazar}(2003)}]{re:lazar03}%
  \BibitemOpen
  \bibfield  {author} {\bibinfo {author} {\bibfnamefont {M.}~\bibnamefont
  {Lazar}},\ }\href@noop {} {\bibfield  {journal} {\bibinfo  {journal} {J.
  Phys. A: Mathematical and General}\ }\textbf {\bibinfo {volume} {36}},\
  \bibinfo {pages} {1415} (\bibinfo {year} {2003})}\BibitemShut {NoStop}%
\bibitem [{\citenamefont {Landau}\ and\ \citenamefont
  {Lifshitz}(1970)}]{re:landau70}%
  \BibitemOpen
  \bibfield  {author} {\bibinfo {author} {\bibfnamefont {L.}~\bibnamefont
  {Landau}}\ and\ \bibinfo {author} {\bibfnamefont {L.}~\bibnamefont
  {Lifshitz}},\ }\href@noop {} {\emph {\bibinfo {title} {Theory of
  elasticity}}}\ (\bibinfo  {publisher} {Pergamon},\ \bibinfo {address} {New
  York},\ \bibinfo {year} {1970})\BibitemShut {NoStop}%
\end{thebibliography}%

\end{document}